\documentclass[aip,pof,reprint]{revtex4-1}

\newcommand{\arcsec}{\ifmmode^{\prime\prime}\else $^{\prime\prime}$\fi}
\newcommand{\arcmin}{\ifmmode^{\prime}\else $^{\prime}$\fi}
\newcommand{\degrees}{\ifmmode^{\circ}\else $^{\circ}$\fi}

\usepackage{amssymb, amsmath}
\usepackage{graphicx}
\usepackage{bm}
\usepackage{color}

\begin{document}

\title{Inverse cascade and symmetry breaking in rapidly-rotating Boussinesq convection}%

\author{B. Favier}%
\email[Corresponding author: ]{favier.benjamin@gmail.com}
\author{L.J. Silvers}
\affiliation{Centre for Mathematical Science, City University London, Northampton Square, London EC1V 0HB, UK}
\author{M.R.E. Proctor}
\affiliation{Department of Applied Mathematics and Theoretical Physics, University of Cambridge, Centre for Mathematical Sciences, Wilberforce Road, Cambridge CB3 0WA, UK}
\date{\today}

\begin{abstract}
In this paper we present numerical simulations of rapidly-rotating Rayleigh-B\'enard convection in the Boussinesq approximation with stress-free boundary conditions.
At moderately low Rossby number and large Rayleigh number, we show that a large-scale depth-invariant flow is formed, reminiscent of the condensate state observed in two-dimensional flows.
We show that the large-scale circulation shares many similarities with the so-called vortex, or slow-mode, of forced rotating turbulence.
Our investigations show that at a fixed rotation rate the large-scale vortex is only observed for a finite range of Rayleigh numbers, as the quasi-two-dimensional nature of the flow disappears at very high Rayleigh numbers.
We observe slow vortex merging events and find a non-local inverse cascade of energy in addition to the regular direct cascade associated with fast small-scale turbulent motions.
Finally, we show that cyclonic structures are dominant in the small-scale turbulent flow and this symmetry breaking persists in the large-scale vortex motion.
\end{abstract}

\maketitle

%
%
\section{Introduction}

A transition between three-dimensional (3D) and two-dimensional (2D) behaviours is observed in many flows due to the action of external forces.
In the particular case of geophysical flows, the combined effects of rotation and density stratification constrain the dynamics due the anisotropic effects of the Coriolis and buoyancy forces.
This leads to two distinct types of motion: large scales are dominated by slow quasi-geostrophic 2D motions whereas small scales are dominated by fast 3D inertia-gravity waves.
The interaction between these small-scale dispersive waves and turbulence is at the core of many models of geophysical flows such as the $\beta$-plane model or the Boussinesq equations for rotating stably-stratified flows.

A generic property of the idealised flows found in geophysical models is their ability to transfer energy to the large-scale slow manifold, providing that rotation or stratification is dominating the dynamics.
For example, inverse cascade of energy from the small forcing scale to the largest available scale has been observed in forced rotating turbulence \cite{smith1999,mininni2009} and forced rotating stably-stratified turbulence \cite{smith2002} in tri-periodic domains.
In these cases, there are no solid boundaries so that the anisotropy and inverse cascade mechanisms originate from the dynamical effects of the body forces only.
Note however that decaying rotating turbulence, even when confined between two horizontal planes\cite{godeferd1999}, does not seem to support an inverse cascade.
The flow in such cases becomes strongly anisotropic with columnar vortices aligned with the rotation axis\cite{cambon1997,staple2008}, but an energy injection mechanism is necessary for the growth of these vortices on larger scales.
It has also been shown possible to gradually shift from a 3D direct cascade to a 2D inverse cascade by reducing one of the spatial dimension of an otherwise tri-periodic system, with or without rotation \cite{smith1996,celani2010}.
Finally, even in fully 3D isotropic turbulence, an inverse cascade of energy can emerge providing that only one type of helical waves is artificially kept in the system\cite{biferale2012}.
The link between these 3D inverse cascade mechanisms in strongly anisotropic flows and the well-known 2D inverse cascade\cite{kraichnan1967}, and the duality between direct and inverse cascades are still being explored\cite{pouquet2013,pouquet2013b}.

Most of the studies about energy cascades in fully developed three-dimensional flows are based on artificial forcings, where the energy is injected arbitrarily at a given scale by a random forcing\cite{smith2002}.
Although the forced approach allows for fine tuning of the energy injection mechanism (in terms of spatial structures\cite{mininni2006} or spatial scales\cite{sukhatme2008,mininni2009} for example), it does not physically represent any realistic instabilities or energy sources.
In the atmosphere, for example, many sources of energy can be invoked to feed the inverse cascade from wave breaking\cite{gage1986} to convective instabilities\cite{vincent1979}.
Rotating convection, where the energy is naturally injected at small scales by the buoyancy force, has been studied in many idealised configurations from homogeneous tri-periodic domains\cite{borue1997b} to the classical Rayleigh-B\'enard configuration.
Although rapidly-rotating Rayleigh-B\'enard convection has been the focus of many experimental and numerical investigations, inverse cascades and condensate states were not observed until recently.
In fully-compressible simulations of rotating convection in a polytropic plane layer, large-scale horizontal flows have been observed when the convection is rapidly-rotating and turbulent\cite{chan2007,kapyla2011}.
Compressible effects modify quantitatively the structure and properties of the large-scale flow, but are not fundamental in order to explain their origin as discussed below.
Using a reduced set of equations, valid in the asymptotic limit of vanishing Rossby number\cite{julien1998}, it has been shown that a similar 2D flow can be sustained even in the Boussinesq approximation \cite{julien2012,rubio2014}.
As yet, however it has not been shown that such an inverse cascade mechanism exists in rapidly-rotating Rayleigh-B\'enard convection at moderate Rossby numbers in models utilising the full set of Boussinesq equations.

In addition to the cascade mechanisms mentioned above, another interesting property of rapidly-rotating fluids is the symmetry breaking between cyclonic structures that rotate in the same direction as the background rotation, and anti-cyclonic structures that rotate oppositely to the background flow.
The dominance of cyclonic vorticity has indeed been observed in rotating turbulence experiments\cite{morize2005} and in many numerical simulations of rotating turbulent flows\cite{bartello1994,bokhoven2008}.
It is known that idealised vortices in a rotating fluid are potentially destabilised by centrifugal instabilities only when their vorticity is negative in the rotating frame\cite{sipp1999,godeferd2001}.
Rotating Rayleigh-B\'enard experiments\cite{vorobieff2002} have also found a correlation between unstable manifolds (as defined by Chong \textit{et al.}\cite{chong1990} in terms of the properties of the local velocity gradient tensor) and anti-cyclonic swirl motions.
However, in all of these studies, the cyclonic dominance was observed in the small-scale turbulent flow only since no inverse cascade mechanism was observed.
In the presence of an inverse cascade, it is not clear how the asymmetry of the small-scale flow will affect the large-scale condensate.
In simulations of rotating compressible convection, different states were obtained, where both cyclones and anti-cyclones can dominate depending on the parameters being considered\cite{mantere2011}.
In the asymptotic model of Julien \textit{et al.}\cite{julien2012,rubio2014}, which is based on an expansion of the Boussinesq equations at low Rossby number, a symmetric state is obtained where both large-scale cyclonic and anti-cyclonic circulations can coexist.
The link between the observed symmetry breaking in the small-scale turbulent flows and large-scale condensate states still remains unclear.

In this paper, we describe the formation of a large-scale, depth-invariant, cyclonic flow in rapidly-rotating Rayleigh-B\'enard convection.
We focus on the Boussinesq regime so that compressible effects, while important in many geophysical and astrophysical contexts, are ignored for now.
The paper is organised as follows.
The equations and the numerical method are described in section \ref{sec:model}.
The choice of parameters for our simulations are discussed in section \ref{sec:param}.
Finally, we discuss our results in section \ref{sec:resbou} and conclude in section \ref{sec:concl}.
%
%
\section{Description of the model\label{sec:model}}
We consider the evolution of a plane-parallel layer of incompressible fluid, bounded above and below by two impenetrable, stress-free walls, a distance $h$ apart.
The geometry of this layer is defined by a Cartesian grid, with $x$ and $y$ corresponding to the horizontal coordinates.
The $z$-axis points vertically upwards.
The layer is rotating about the $z$-axis with a constant angular velocity $\bm{\Omega}=\Omega\hat{\bm{z}}$ and gravity is pointing downwards $\bm{g}=-g\hat{\bm{z}}$.
The horizontal size of the fluid domain is defined by the aspect ratio $\lambda$ so that the fluid occupies the domain $0<z<h$, $0<x<\lambda h$ and $0<y<\lambda h$.
The kinematic viscosity, $\nu$, and thermal conductivity, $\kappa$, are assumed to be constant.

In the Boussinesq approximation, using the thermal diffusion time in the vertical as a unit of time and the depth of the layer $h$ as a unit of length, the dimensionless equations are
\begin{equation}
\label{eq:momentum}
\frac{\partial\bm{u}}{\partial t}=-\bm{u}\cdot\nabla\bm{u}-\nabla p-\sigma\sqrt{Ta}\bm{e}_z\times\bm{u}+\sigma Ra \theta \bm{e}_z+\sigma\nabla^2\bm{u} \ ,
\end{equation}
\begin{equation}
\frac{\partial\theta}{\partial t}=-\bm{u}\cdot\nabla\theta+w+\nabla^2\theta \ ,
\end{equation}
\begin{equation}
\label{eq:div}
\nabla\cdot\bm{u}=0 \ ,
\end{equation}
where $\bm{u}=\left(u, v, w\right)$ is the velocity, $p$ is the pressure and $\theta$ is the fluctuating temperature with respect to a linear background.
$Ra$ is the Rayleigh number, $Ta$ is the Taylor number and $\sigma$ is the Prandtl number defined in the usual way by
\begin{equation}
Ra=\frac{g\alpha\Delta Th^3}{\nu\kappa} \ , \quad Ta=\frac{4\Omega^2h^4}{\nu^2} \quad \textrm{and} \quad \sigma=\frac{\nu}{\kappa} \ .
\end{equation}
For simplicity, the Prandtl number is fixed to be unity throughout the paper.
These dimensionless quantities involve $g$ the constant gravitational acceleration, $\alpha$ the coefficient of thermal expansion and $\Delta T$ the temperature difference between the two horizontal plates.

In the horizontal directions, all variables are assumed to be periodic.
The upper and lower boundaries are assumed to be impermeable and stress-free, which implies that $\partial_zu = \partial_z v = w = 0$ at $z=0$
and $z=1$.
The thermal boundary conditions at these surfaces correspond to fixing $\theta=0$ at $z=0$ and $z=1$.

\setlength{\tabcolsep}{5.5pt}
\begin{table*}
\centering
\caption{Summary of the parameters considered in this study. $Ta$ is the Taylor number, $Ra$ is the Rayleigh number, $Ra_c$ is the critical Rayleigh number, $Ro$ is the input Rossby number as defined by equation \eqref{eq:ro}, $\tilde{Ra}$ is a scaled Rayleigh number as defined by equation \eqref{eq:ratilde}, $\lambda$ is the horizontal aspect ratio, $k_c$ is the most unstable wave number predicted by linear theory \cite{chan61}, $U_{\textrm{rms}}$ is the root mean square velocity, $L_0$ is the horizontal integral scale, $Re$ is the Reynolds number, $Ro_L$ is the large-scale Rossby number and $Ro_{\omega}$ is the small-scale Rossby number. The Prandtl number $\sigma$ is fixed to be one for all simulations. The output quantities computed from the simulations are obtained by removing the vortex mode as defined by equations~\eqref{eq:upr}-\eqref{eq:vpr}. LSC stands for Large-Scale Circulation.\label{tab:one}}
\begin{tabular}{@{}c|cccccccc|ccccc|c@{}}
\hline
\hline
Case & $Ta$ & $Ra$ & $Ra/Ra_c$ & $Ro$ & $\tilde{Ra}$ & $\lambda$ & $2\pi/k_c$ & $Nx \! \times \! Ny \! \times \! Nz$ & $U_{\textrm{rms}}$ & $L_0$ & $Re$ & $Ro_L$ & $Ro_{\omega}$ & LSC\\
\hline
A1 & $10^6$ & $5\times10^5$ & $5.8$ & $0.71$ & $50$ & $4$ & $0.48$ & $240^2\times 120$ & $124$ & $0.19$ & $23.6$ & $0.65$ & $1.82$ & No\\ 
A2 & $10^6$ & $2\times10^6$ & $23.0$ & $1.4$ & $200$ & $4$ & $0.48$ & $480^2\times 120$ & $291$ & $0.18$ & $52.3$ & $1.62$ & $5.17$ & No\\ 
A3 & $10^6$ & $10^7$ & $115.0$ & $3.2$ & $10^3$ & $6$ & $0.48$ & $720^2\times 240$ & $704$ & $0.15$ & $105.6$ & $4.69$ & $15.4$ & No\\[0.2cm]
B1 & $10^7$ & $10^6$ & $2.5$ & $0.32$ & $21.5$ & $4$ & $0.33$ & $240^2\times 120$ & $100$ & $0.13$ & $13.4$ & $0.24$ & $0.63$ & No\\ 
B2 & $10^7$ & $2\times10^6$ & $5.0$ & $0.45$ & $43.1$ & $4$ & $0.33$ & $240^2\times 120$ & $201$ & $0.14$ & $28.2$ & $0.46$ & $1.33$ & No\\ 
B3 & $10^7$ & $5\times10^6$ & $12.4$ & $0.71$ & $107.7$ & $4$ & $0.33$ & $360^2\times 180$ & $380$ & $0.16$ & $60.9$ & $0.75$ & $2.83$ & No\\[0.2cm] 
C1 & $10^8$ & $2\times10^6$ & $1.1$ & $0.14$ & $9.3$ & $6$ & $0.22$ & $240^2\times 96$ & $14.3$ & $0.05$ & $0.67$ & $0.03$ & $0.04$ & No\\ 
C2 & $10^8$ & $4\times10^6$ & $2.1$ & $0.2$ & $18.6$ & $6$ & $0.22$ & $360^2\times 120$ & $123$ & $0.09$ & $11.2$ & $0.14$ & $0.35$ & No\\ 
C3 & $10^8$ & $10^7$ & $5.3$ & $0.32$ & $46.4$ & $6$ & $0.22$ & $480^2\times 120$ & $375$ & $0.12$ & $43.8$ & $0.32$ & $1.13$ & Yes\\ 
C4 & $10^8$ & $2\times10^7$ & $10.7$ & $0.45$ & $92.8$ & $4$ & $0.22$ & $360^2\times 180$ & $625$ & $0.11$ & $70.6$ & $0.55$ & $2.1$ & Yes\\ 
C5 & $10^8$ & $5\times10^7$ & $26.7$ & $0.71$ & $232.1$ & $5$ & $0.22$ & $480^2\times 240$ & $1072$ & $0.11$ & $121.2$ & $0.95$ & $4.1$ & No\\ 
C6 & $10^8$ & $10^8$ & $53.4$ & $1$ & $464.2$ & $4$ & $0.22$ & $480^3$ & $1569$ & $0.12$ & $193.0$ & $1.28$ & $6.76$ & No\\[0.2cm] 
D1 & $10^9$ & $10^7$ & $1.2$ & $0.1$ & $10$ & $3$ & $0.15$ & $240^2\times 120$ & $38$ & $0.05$ & $2.0$ & $0.02$ & $0.05$ & No\\ 
D2 & $10^9$ & $2\times10^7$ & $2.3$ & $0.14$ & $20$ & $3$ & $0.15$ & $240^2\times 120$ & $214$ & $0.06$ & $13.9$ & $0.1$ & $0.28$ & No\\ 
D3 & $10^9$ & $5\times10^7$ & $5.8$ & $0.22$ & $50$ & $3$ & $0.15$ & $360^2\times 180$ & $658$ & $0.09$ & $61.2$ & $0.22$ & $0.94$ & Yes\\ 
D4 & $10^9$ & $10^8$ & $11.5$ & $0.32$ & $100$ & $3$ & $0.15$ & $360^2\times 180$ & $1111$ & $0.1$ & $111.1$ & $0.35$ & $1.7$ & Yes\\ 
D5 & $10^9$ & $3\times10^8$ & $34.5$ & $0.55$ & $300$ & $3$ & $0.15$ & $480^2\times 240$ & $2241$ & $0.1$ & $232.6$ & $0.68$ & $4.1$ & Yes\\ 
D6 & $10^9$ & $5\times10^8$ & $57.5$ & $0.71$ & $500$ & $2$ & $0.15$ & $480^3$ & $2997$ & $0.09$ & $260.0$ & $1.1$ & $6.0$ & No\\[0.2cm] 
E1 & $10^{10}$ & $5\times10^7$ & $1.2$ & $0.07$ & $10.8$ & $4$ & $0.1$ & $480^2\times 120$ & $75$ & $0.02$ & $1.8$ & $0.03$ & $0.05$ & No\\ 
E2 & $10^{10}$ & $10^8$ & $5.0$ & $0.14$ & $43.1$ & $4$ & $0.1$ & $600^2\times 120$ & $354$ & $0.04$ & $15.3$ & $0.08$ & $0.21$ & Yes\\ 
E3 & $10^{10}$ & $2\times10^8$ & $9.9$ & $0.14$ & $86.2$ & $4$ & $0.1$ & $600^2\times 180$ & $908$ & $0.06$ & $58.0$ & $0.14$ & $0.56$ & Yes\\ 
E4 & $10^{10}$ & $5\times10^8$ & $12.4$ & $0.22$ & $107.7$ & $4$ & $0.1$ & $720^2\times 240$ & $1969$ & $0.07$ & $133.3$ & $0.29$ & $1.39$ & Yes\\ 
E5 & $10^{10}$ & $10^9$ & $24.8$ & $0.32$ & $215.4$ & $2$ & $0.1$ & $480^3$ & $3302$ & $0.07$ & $240.1$ & $0.45$ & $2.58$ & Yes\\[0.2cm] 
F1 & $10^{12}$ & $10^9$ & $1.2$ & $0.03$ & $10$ & $1$ & $0.05$ & $240^2\times 120$ & $122$ & $0.01$ & $1.5$ & $0.01$ & $0.02$ & No\\ 
F2 & $10^{12}$ & $3\times10^9$ & $3.5$ & $0.05$ & $30$ & $1$ & $0.05$ & $480^2\times 240$ & $1297$ & $0.03$ & $43.3$ & $0.04$ & $0.16$ & Yes\\ 
F3 & $10^{12}$ & $10^{10}$ & $11.5$ & $0.1$ & $100$ & $1$ & $0.05$ & $480^3$ & $4790$ & $0.05$ & $215.5$ & $0.11$ & $0.67$ & Yes\\ 
\hline
\hline
\end{tabular}
\end{table*}

We solve equations \eqref{eq:momentum}-\eqref{eq:div} using a well-tested mixed pseudo-spectral and finite difference method.
The original numerical method is based on the compressible code described in Matthews \textit{et al.}\cite{matt95} and used in many subsequent papers \cite{bushby08,silvers2009,favier2012,favier2012b}.
We have updated the code to solve the Boussinesq equations given above in the same Cartesian geometry.
In order to ensure the incompressibility condition defined by equation \eqref{eq:div}, the velocity field is decomposed using a poloidal-toroidal formulation
\begin{equation}
\label{eq:poltor}
\bm{u}=\nabla\times\nabla\times\left(S\hat{\bm{z}}\right)+\nabla\times\left(T\hat{\bm{z}}\right) \ ,
\end{equation}
where $T$ is the toroidal component, $S$ is the poloidal component and $\hat{\bm{z}}$ is the unit vector in the vertical direction.
Each of these scalar quantities is projected onto a horizontal Fourier basis
\begin{align}
S(x,y,z) & = \sum\hat{S}(k_x,k_y,z)e^{ik_xx}e^{ik_yy} \ , \\
T(x,y,z) & = \sum\hat{T}(k_x,k_y,z)e^{ik_xx}e^{ik_yy} \ ,
\end{align}
where $k_x$ and $k_y$ are discrete horizontal that are wave numbers multiple of $2\pi/\lambda$.
For each value of the horizontal wave numbers, the vertical functions $\hat{S}(k_x,k_y,z)$ and $\hat{T}(k_x,k_y,z)$ are represented by their discretized values on a non-uniform vertical grid between $z=0$ and $z=1$.
The vertical derivatives are computed using fourth order finite-differences and boundary conditions are imposed using ghost points at the top and bottom boundaries.
The grid is denser close to the boundaries in order to appropriately resolve thermal boundary layers.
The time-stepping is performed using a semi-implicit second order Crank-Nicholson scheme for the linear dissipative terms and a third-order explicit Adams-Bashforth for the nonlinear advection and linear Coriolis terms.
The solution is de-aliased in the horizontal directions using the $2/3$ rule.
The spectral order in the horizontal direction is varied from $240$ up to $720$ depending on the aspect ratio and parameters considered, the number of grid points in the vertical direction is varied from $96$ up to $480$ and the code is parallelised in the vertical direction using MPI.

The numerical approach that we are utilising for this work is similar to the one used by Cattaneo \textit{et al.}\cite{cattaneo2003} (apart from the use of finite differences in the vertical direction instead of Fourier modes in their case).
In addition to testing our new version of the code against the results of Cattaneo \textit{et al.}, we have also successfully compared our numerical scheme against the spectral element code {\sc Nek5000}\cite{Fischer2007,nek5000}.

%
%
\section{Parameters \label{sec:param}}

The aim of this paper is to find the parameters for which a large-scale circulation driven by an inverse cascade is observed, and discuss the properties of this large-scale flow as a function of the key dimensionless parameters of this problem.
It is known from previous compressible studies \cite{chan2007,kapyla2011,mantere2011} and from the recent works using asymptotic model equations \cite{julien1998,julien2012,rubio2014} that a large-scale circulation only appears if the flow is sufficiently constrained by rotation (low Rossby number regime) and sufficiently turbulent (large Reynolds number regime).
Therefore, this is the regime this paper is focusing on.

We define here the large-scale (\textit{i.e.} based on the depth of the fluid layer) Rossby number as
\begin{equation}
\label{eq:ro}
Ro=\sqrt{\frac{Ra}{\sigma Ta}}
\end{equation}
where $Ra$, $\sigma$ and $Ta$ are the Rayleigh, Prandtl and Taylor numbers respectively.
In order to estimate the level of turbulent activity \textit{a priori}, we use the ratio between the Rayleigh number and the critical Rayleigh number (which depends on the Taylor number) in order to describe how far we are from the onset of thermal convection.
Finally, it is well known from linear theory\cite{chan61} that the ratio between the characteristic horizontal length scale of the flow and the depth of the layer scales as $Ta^{-1/6}$.
In order to take into account for this strong variability in the horizontal length-scale of the flow as the Taylor number is varied, we use the scaled Rayleigh number
\begin{equation}
\label{eq:ratilde}
\tilde{Ra}=Ra \ Ta^{-2/3} \ ,
\end{equation}
which was used in the asymptotic model of Julien \textit{et al.}\cite{julien2012}.

In this investigation, we vary the Taylor number from $10^6$ up to $10^{12}$ and the Rayleigh number from $10^{5}$ up to $10^{10}$.
For each Taylor number, the Rayleigh number considered are from just above the critical threshold up to 1000 times this value.
This choice of parameters implies that the Rossby number, as defined by equation~\eqref{eq:ro}, is varying between $0.03$ and $3.2$ and the scaled Rayleigh number $\tilde{Ra}$, as defined by equation~\eqref{eq:ratilde}, is varying from $10$ to $10^3$.
Our complete set of parameters for each of the simulations discussed in this paper is summarized in Table~\ref{tab:one}.

In order to describe our results, we calculate the following quantities.
We first compute the root mean square velocity $U_{\textrm{rms}}=\sqrt{\left<\bm{u}^2\right>}$ in the early stage of the nonlinear evolution (just after the system reaches a quasi-steady equilibrium and before an eventual steady increase of the kinetic energy), where $\left<.\right>$ denotes the volume average.
A typical horizontal length-scale of the flow is then estimated by computing the longitudinal correlation length-scale as
\begin{equation}
L_0=\frac12\int_0^{\lambda}\frac{\left<u_i(\bm{x}+l\bm{e}_i)u_i(\bm{x})\right>}{\left<u_i(\bm{x})u_i(\bm{x})\right>}\textrm{d}l
\end{equation}
where $u_i$ and $\bm{e}_i$ are the velocity component and the unit vector in one of the horizontal directions.
We then define the local Reynolds and Rossby numbers as
\begin{equation}
\label{eq:re}
Re=U_{\textrm{rms}}L_0 \quad \textrm{and} \quad Ro_L=\frac{U_{\textrm{rms}}}{\sqrt{Ta} L_0} \ .
\end{equation}
Note that the Reynolds number does not depend explicitly on viscosity since we use the thermal or viscous diffusion time as a unit of time.
In keeping with previous publications on rotating turbulence\cite{bartello1994,cambon1997}, we also define a small-scale Rossby number based on the vertical component of the vorticity as
\begin{equation}
Ro_{\omega}=\frac{\omega_{\textrm{rms}}}{\sqrt{Ta}} \ ,
\end{equation}
where $\omega_{\textrm{rms}}$ is the root mean square value of the vorticity $\bm{\omega}=\nabla\times\bm{u}$.
The values of these output parameters for each case are given in the right part of Table~\ref{tab:one}.
Note that these dimensionless parameters correspond to the flow associated with the 3D mode only, which is defined in the next section by equations~\eqref{eq:upr}-\eqref{eq:wpr}.

%
%
\section{Results\label{sec:resbou}}

\subsection{Growth of the 2D mode}

\begin{figure}
   \hspace{-0.8cm}
   \includegraphics[width=80mm]{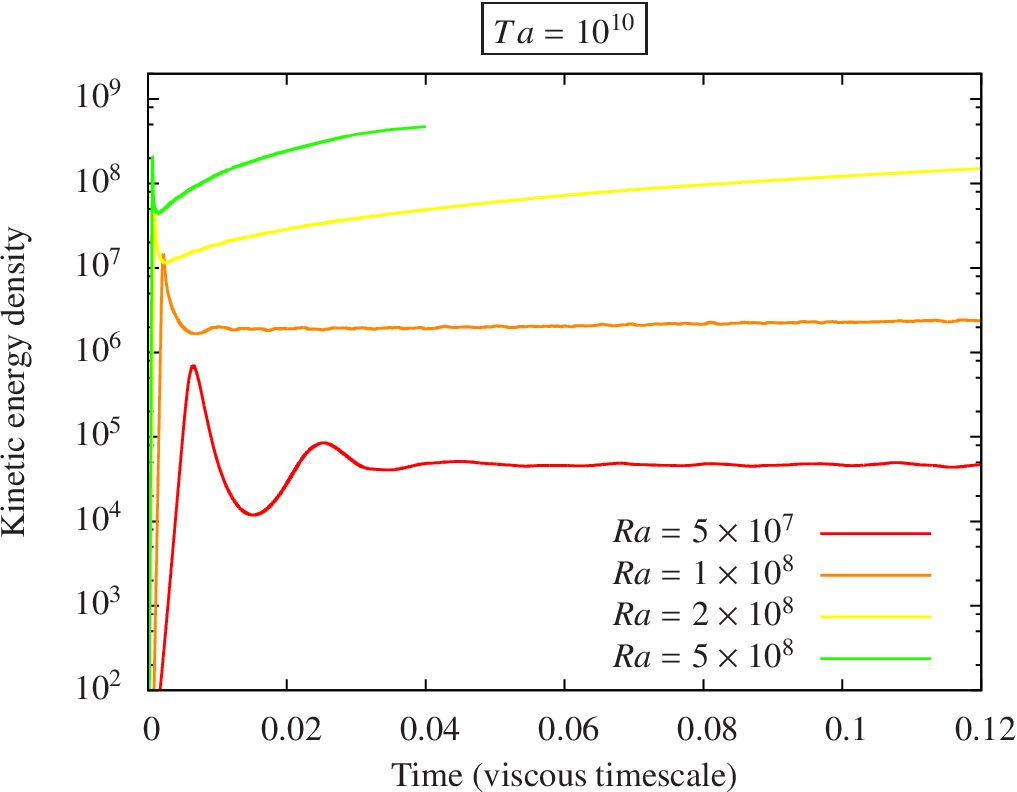}
   \caption{Time evolution of the kinetic energy density for cases $E1$ to $E4$ in Table~\ref{tab:one}.\label{fig:kin}}
\end{figure}

\begin{figure*}
   \includegraphics[height=50mm]{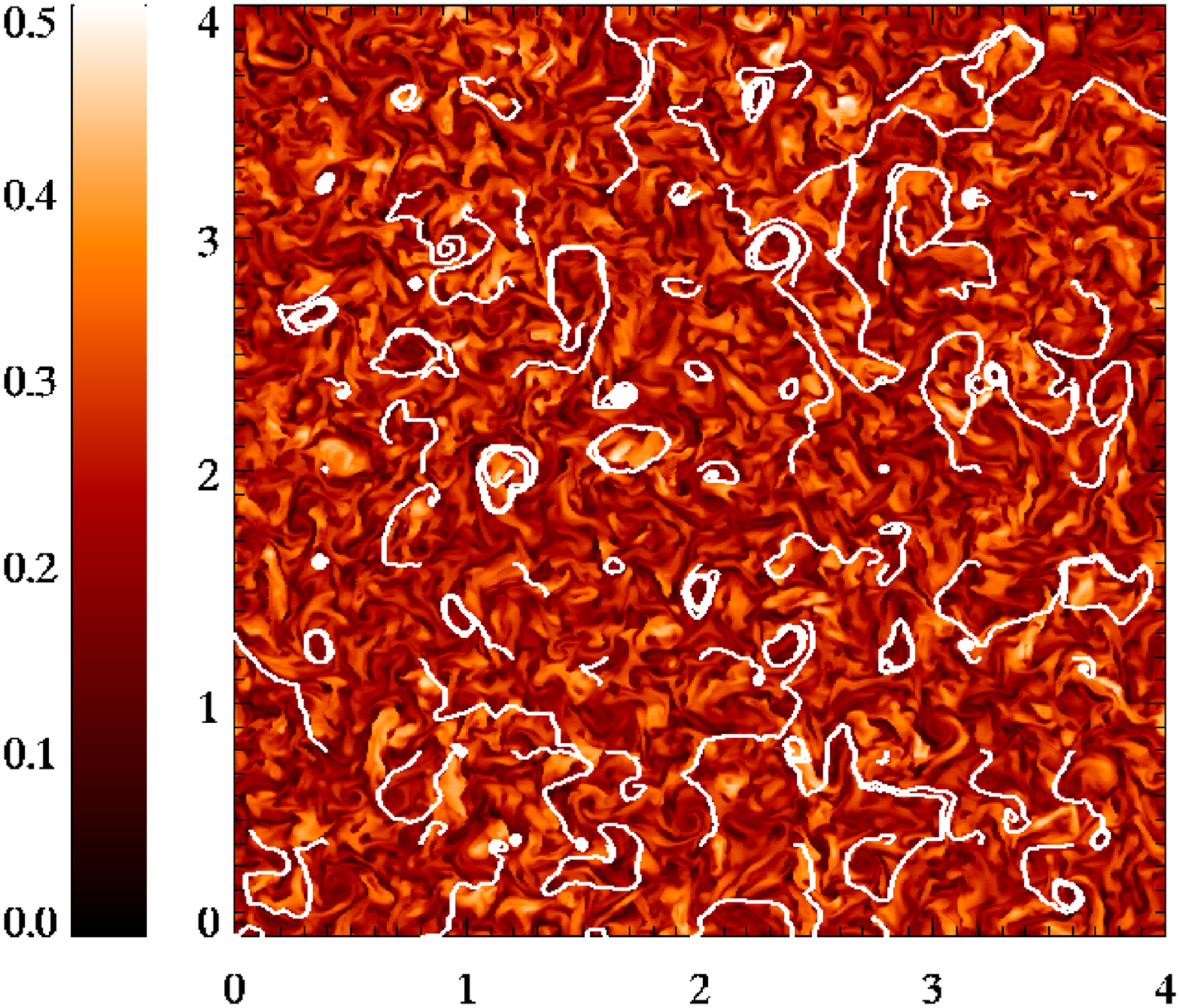}
   \includegraphics[height=50mm]{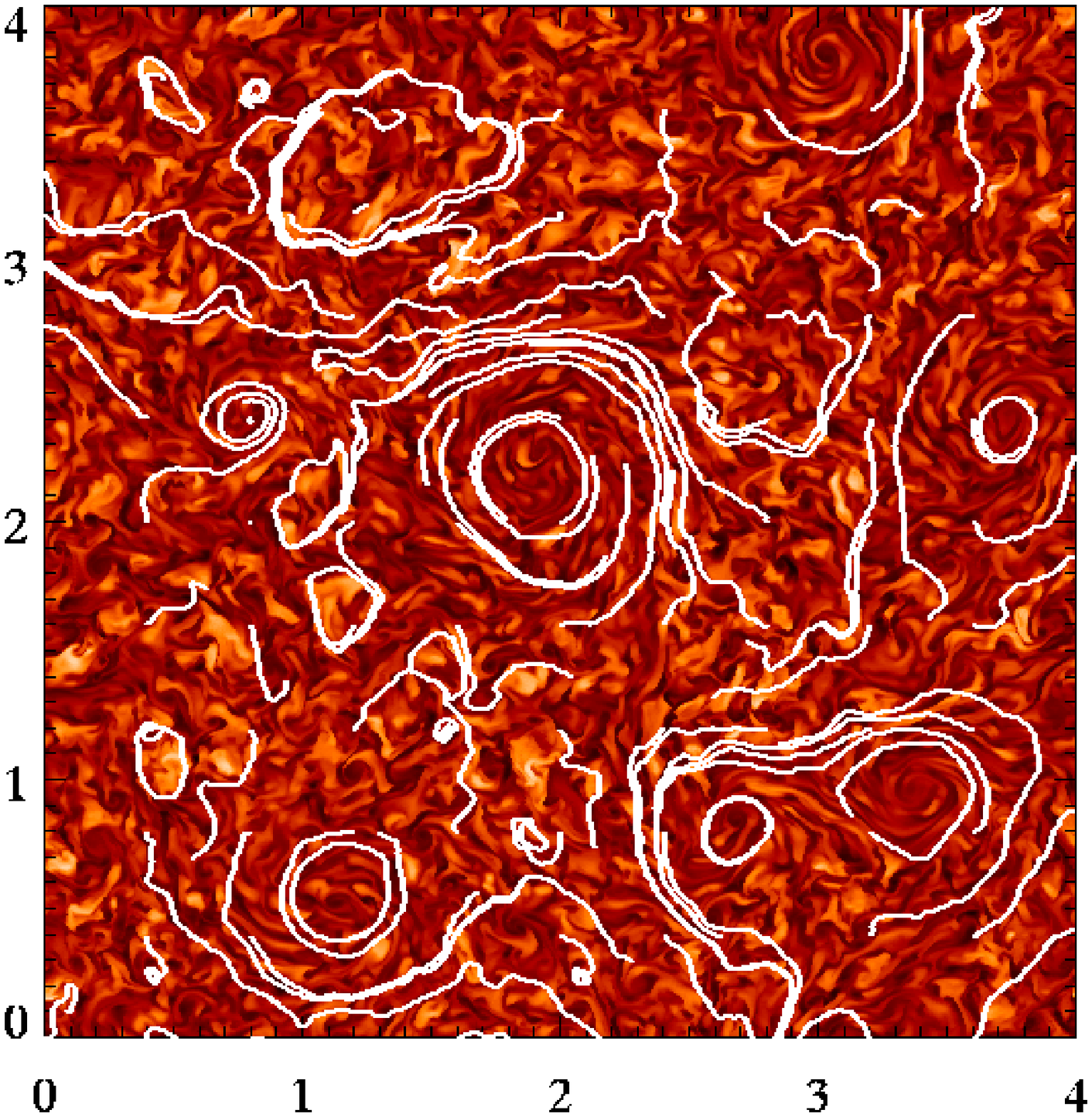}
   \includegraphics[height=50mm]{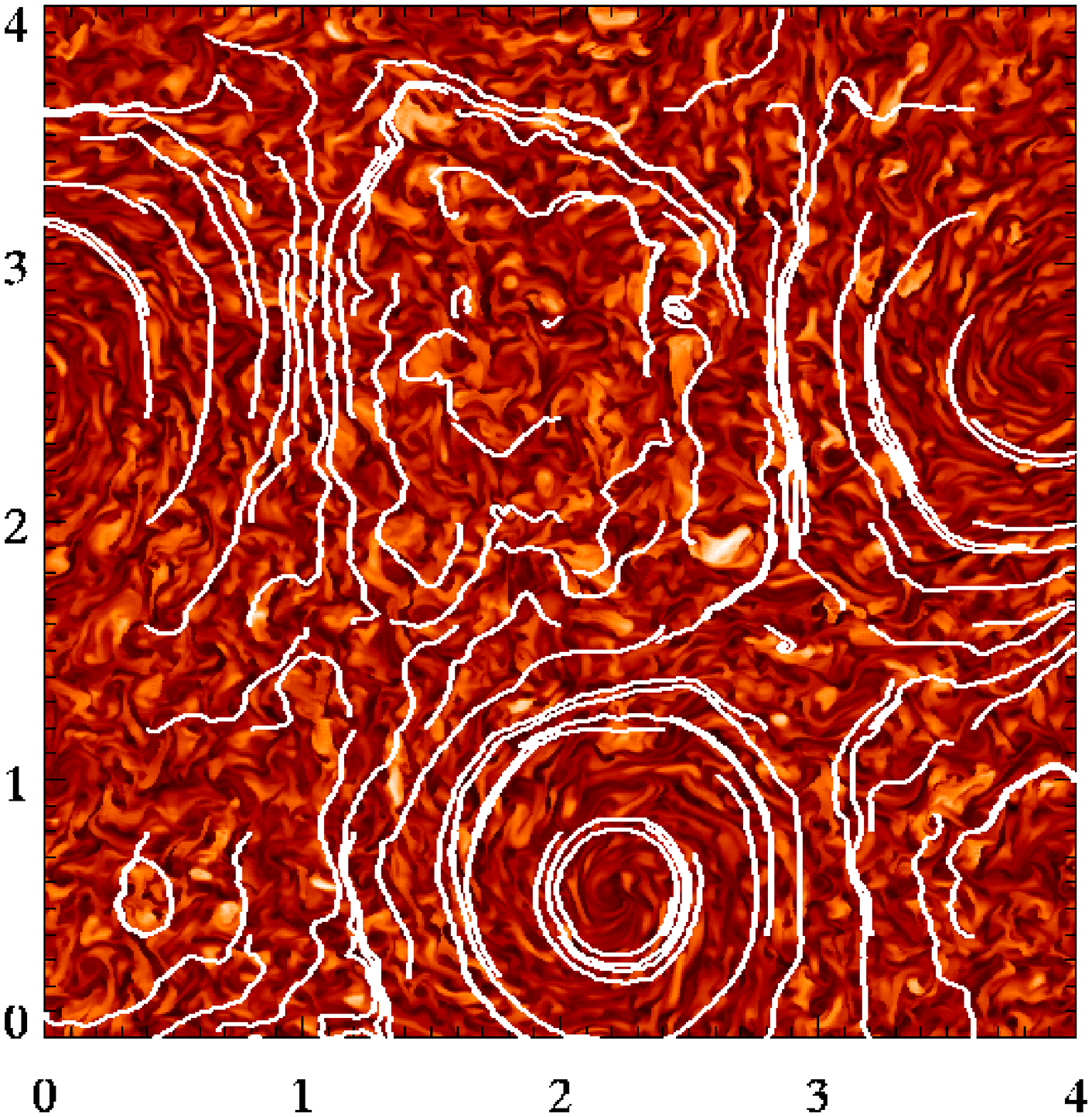}\\
   \vspace{2.4mm}
   \hspace{-4mm}
   \includegraphics[height=50mm]{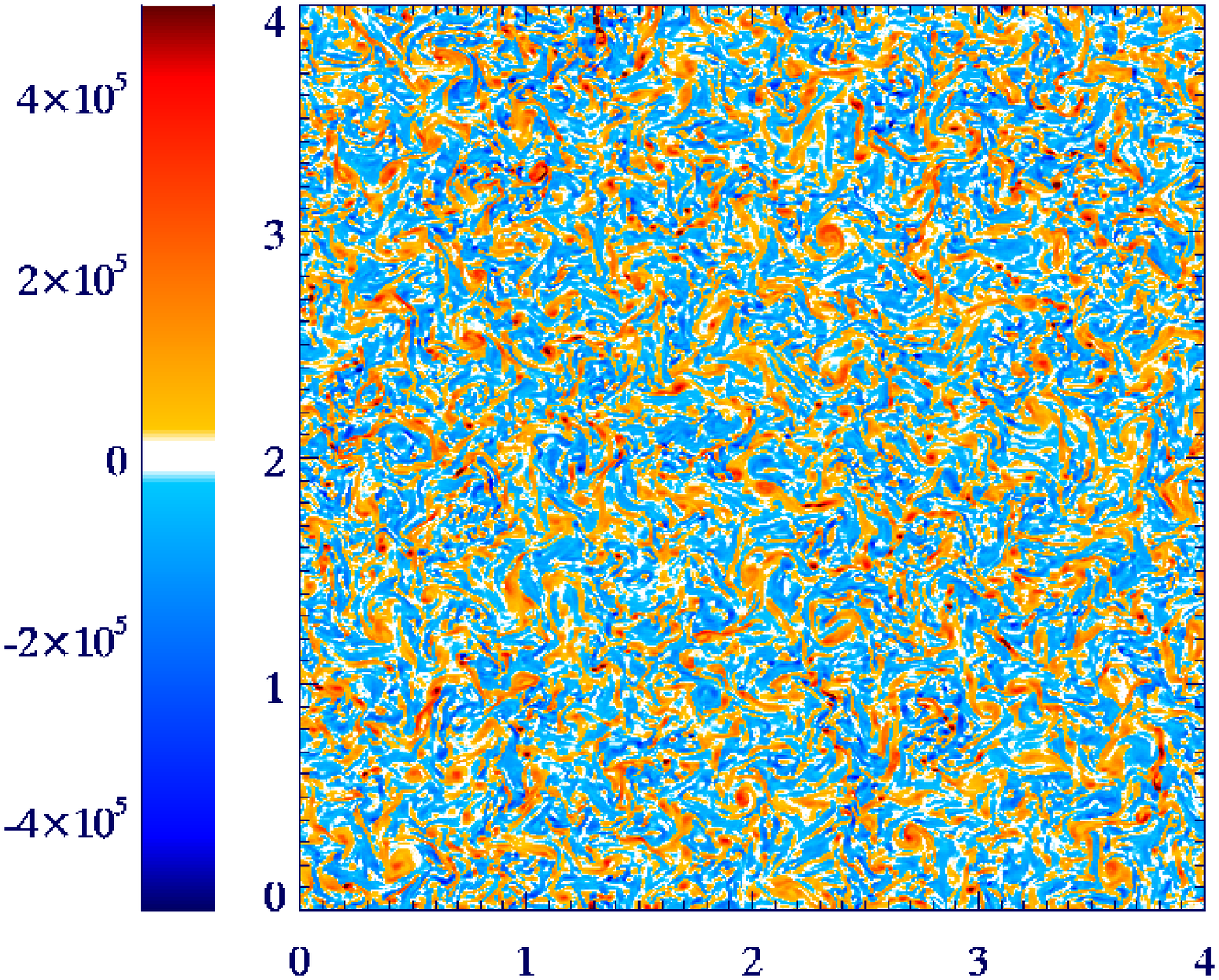}
   \includegraphics[height=50mm]{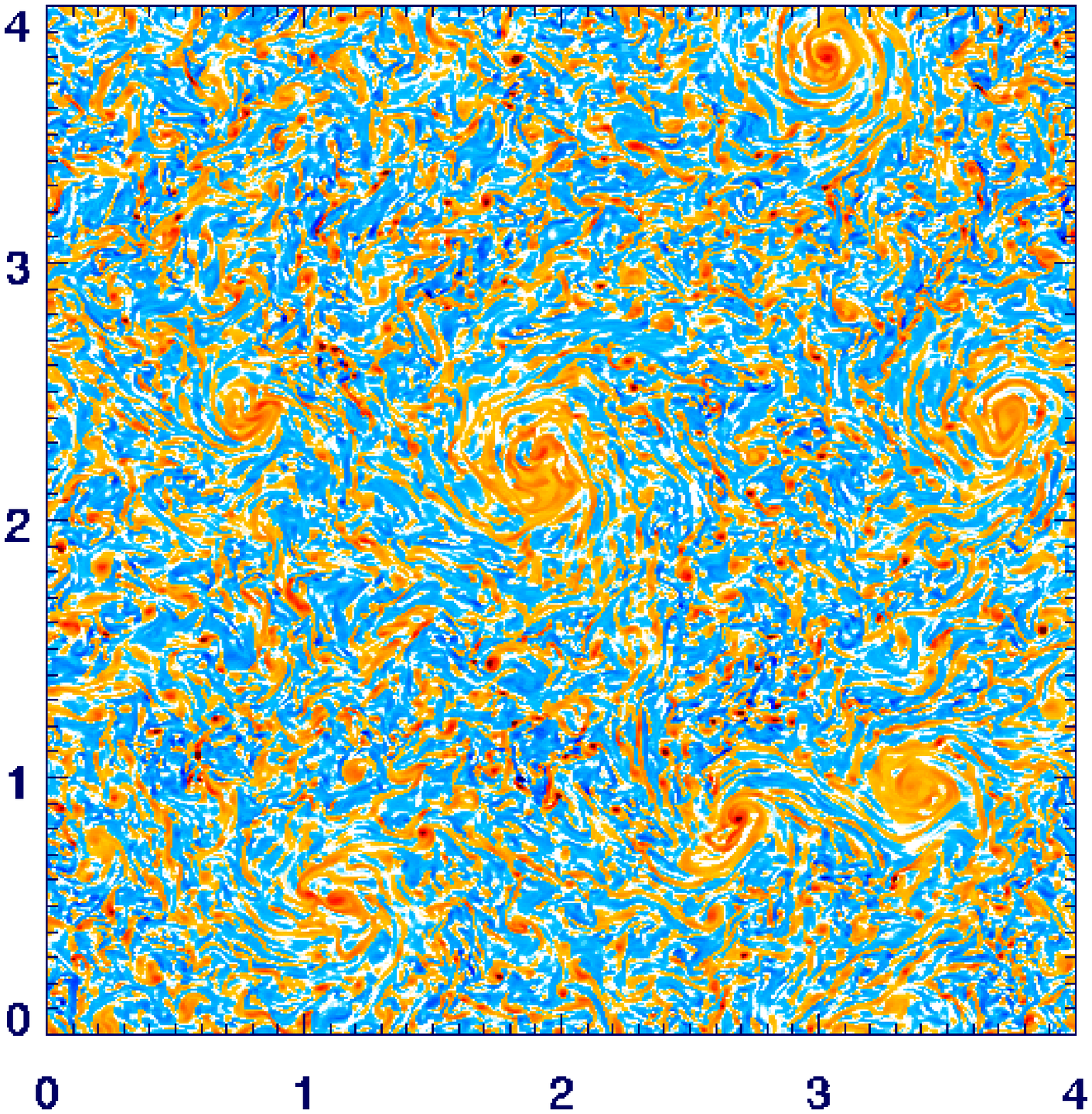}
   \includegraphics[height=50mm]{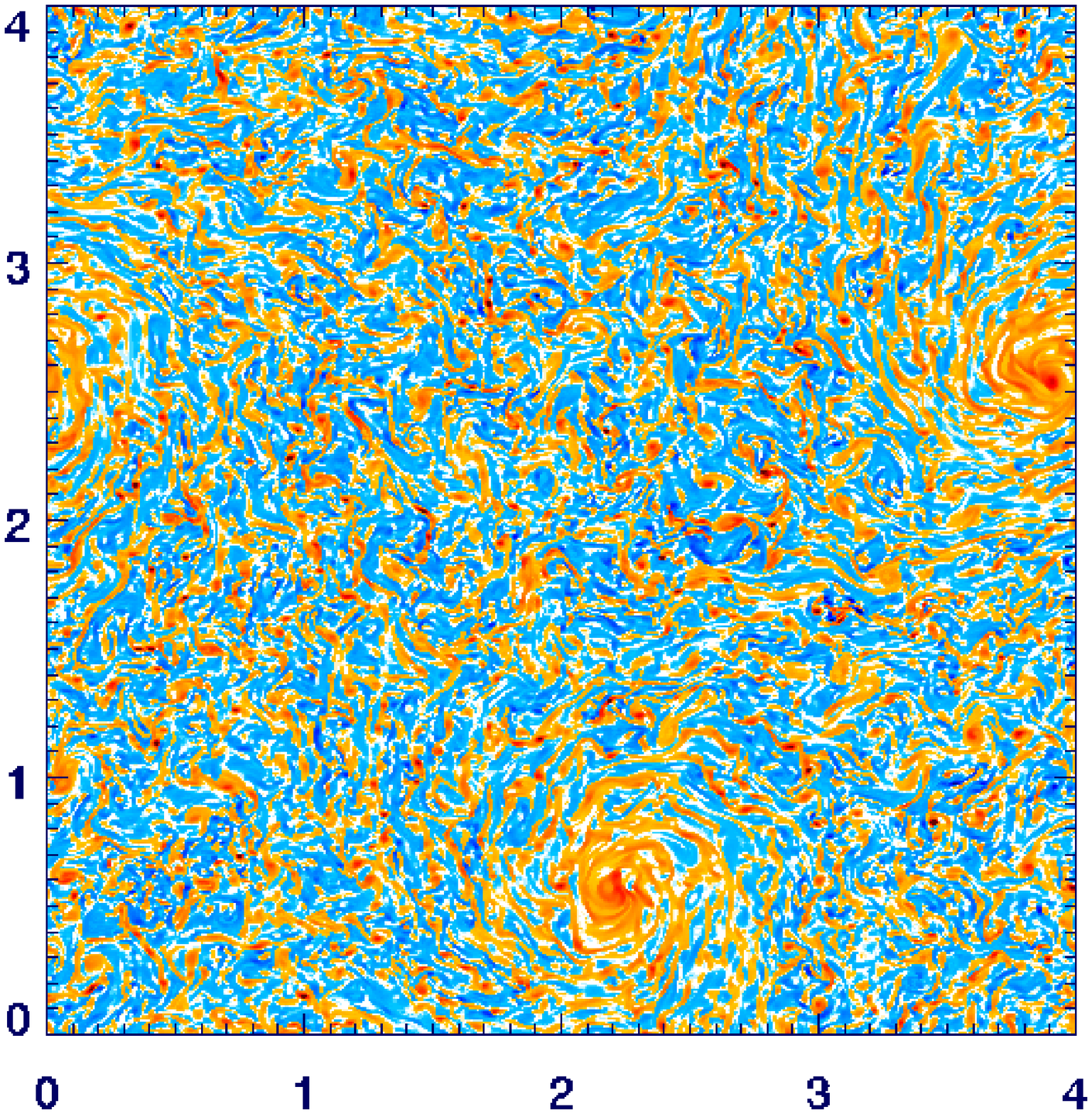}
   \caption{Top line: temperature fluctuations in a horizontal plane located at $z=0.95$, close to the upper boundary. Bright/yellow colours correspond to positive temperature fluctuations whereas dark/red colors correspond to negative temperature fluctuations. We also plot horizontal streamlines as white lines in order to emphasis the large-scale circulation. Bottom line: vertical vorticity in the middle of the layer $z=0.5$. Red corresponds to positive cyclonic vorticity whereas blue corresponds to negative anti-cyclonic vorticity. Time is increasing from left to right and snapshots are taken at $t=0.001$, $t=0.01$ and $t=0.02$. The figures on the left correspond to the very beginning of the nonlinear phase and each plot is then separated by approximately two hundred turnover times.\label{fig:temp}}
\end{figure*}

In this section, we discuss a particular set of simulations where the Taylor number is fixed to be $Ta=10^{10}$ (cases $E1$ to $E4$ in Table~\ref{tab:one}).
The kinetic energy density as a function of time for these cases is shown in Figure~\ref{fig:kin}.
This shows an initial linear phase where the kinetic energy grows exponentially from a layer at rest with small temperature perturbations.
As expected, the growth rate of the instability increases as the Rayleigh number is increased.
After the nonlinear saturation, the kinetic energy eventually evolves towards its equilibrium value.
For the simulation with the smallest Rayleigh number, $E1$, the kinetic energy remains close to its quasi-stationary value for a complete viscous timescale.
For larger Rayleigh numbers however (simulations $E3$ and $E4$ for example), the kinetic energy gradually increases with time.
After many turnover times, the kinetic energy eventually saturates to a much larger value than what is observed for the initial nonlinear saturation.
During the slow growth phase of the kinetic energy, both total viscous dissipation and buoyancy work (not shown) are found to remain fairly constant.
As the kinetic energy saturates, and buoyancy balances viscous forces, we observe a slight decrease in both of these quantities.
This corresponds to an overall decrease of the Nusselt number and heat flux through the layer, as discussed in more detail by Guervilly \textit{et al.}\cite{guervilly2014}.

Figure~\ref{fig:temp} shows a horizontal slice across the domain for case $E4$ at times $t=0.001$, $t=0.01$ and $t=0.02$.
We show the the temperature fluctuations in the plane $z=0.95$, close to the upper boundary, and the vertical vorticity at the middle of the layer, $z=0.5$.
At $t=0.001$, small-scale convective cells are clearly identifiable.
For this particular value of the Taylor number, the horizontal scale of maximum growth rate predicted by linear theory\cite{chan61} is approximately $2\pi/k_c=0.1$ (which we have verified numerically).
Given this horizontal scale, the aspect ratio of $\lambda=4$ provides an initial clear scale separation.
As time increases, we observe the generation of vortices evolving on a horizontal scale much larger than the one of the background small-scale convective flow.
This is also seen by looking at the horizontal streamlines in the top panel of Figure~\ref{fig:temp}.
Note that the streamlines associated with positive cyclonic vertical vorticity are smooth when compared to the streamlines associated with the anti-cyclonic regions, which will be further discussed in section \ref{sec:cycl}.
For all the simulations considered in this paper, the large-scale circulation is growing in amplitude on a timescale much larger than the turnover time associated with the small-scale convective flow.

As discussed by Julien \textit{et al.}\cite{julien2012}, the large-scale circulation shown in Figure~\ref{fig:temp} is better described as a depth-independent 2D flow, which they called barotropic circulation.
This decomposition between slow quasi-geostrophic and fast 3D flows has been used for a long time in order to describe rotating and stratified homogeneous turbulence\cite{bart1995,ca2001,smith2002,bourouiba2012}.
In the case of a vertically rotating fluid for example, the dispersion relation of inertial waves is $\omega=2\Omega k_z/k$ where $\omega$ is the wave frequency, $k_z$ is the wave number in the vertical direction and $k$ is the modulus of the wave vector.
The so-called slow vortex mode corresponds to purely horizontal wave vectors (in the case of a vertical rotation axis) such that the inertial wave frequency is zero.
In our case, inertial waves are present due to the restoring Coriolis force, but internal gravity waves are not supported since the stratification is unstable, leading instead to convective eddies.
In addition, our system is inhomogeneous in the vertical direction due the presence of the horizontal solid boundaries.
For these reasons, we cannot explicitly use the classical wave-vortex decomposition, but we introduce a very similar flow decomposition, following Julien \textit{et al.}\cite{julien2012}.
Note also that it has been shown that in the case of wave turbulence in a rotating channel, the 2D depth-invariant mode needs to be considered independently, since the group velocity of inertial modes vanishes in this case, which is inconsistent with classical wave turbulence theory based on dispersive waves\cite{scott2014}.

\begin{figure}
   \hspace{-0.8cm}
   \includegraphics[width=80mm]{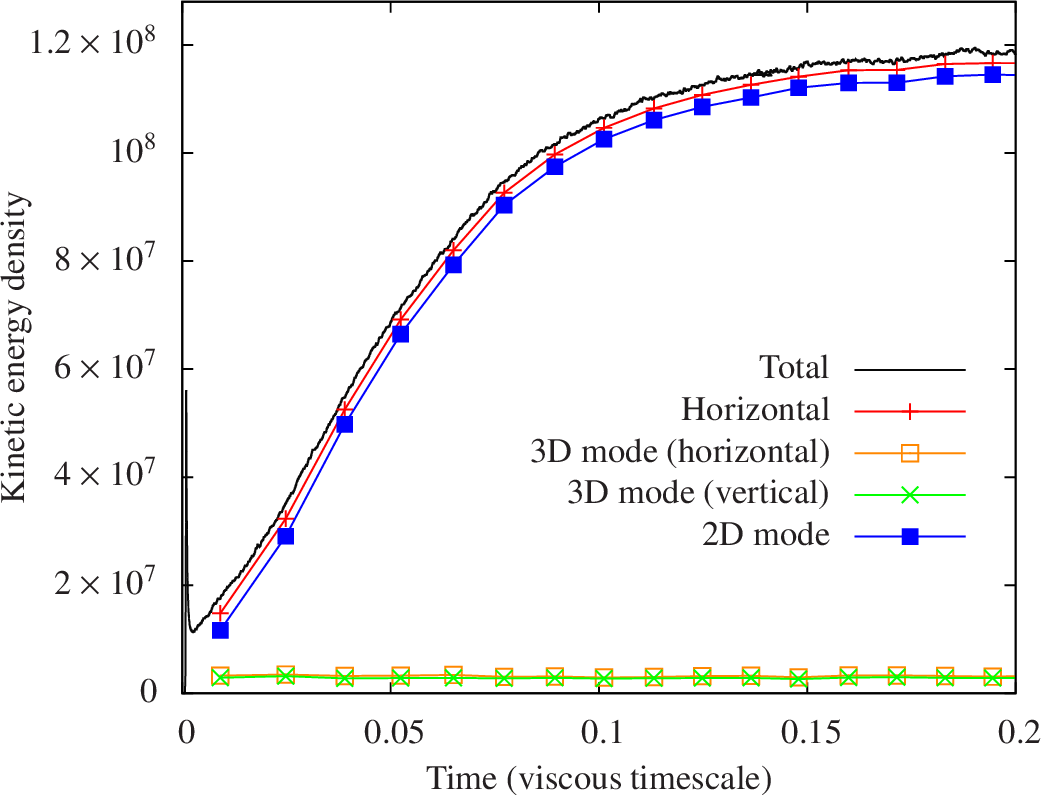}
   \caption{Time evolution of the kinetic energy for simulation $E3$. The total kinetic energy (straight black line) is decomposed between horizontal, 2D (equation \eqref{eq:barot}), 3D horizontal (equation \eqref{eq:baroc}) and 3D vertical components (equation \eqref{eq:barocv}).\label{fig:baro}}
\end{figure}

In the presence of the two solid boundaries at $z=0$ and $z=1$, we define the depth-averaged 2D horizontal flow, subsequently called the 2D mode, as
\begin{align}
\label{eq:vortex}
\left<u\right>_z\!(x,y) & = \int_0^1 u(x,y,z) \, \textrm{d}z \\
\label{eq:vortex2}
\left<v\right>_z\!(x,y) & = \int_0^1 v(x,y,z) \, \textrm{d}z \ ,
\end{align}
where $u$ and $v$ are the velocity component in the $x$ and $y$ directions respectively, and the fast 3D fluctuations, subsequently called the 3D mode, as
\begin{align}
\label{eq:upr}
u'(x,y,z) & = u(x,y,z)-\left<u\right>_z(x,y) \\
\label{eq:vpr}
v'(x,y,z) & = v(x,y,z)-\left<v\right>_z(x,y) \\
\label{eq:wpr}
w'(x,y,z) & = w(x,y,z) \ .
\end{align}
Note that the vertical average of the vertical velocity is zero due to mass conservation, so that the vertical velocity only appears in the 3D mode.
We can then define the (purely horizontal) kinetic energy associated with the slow 2D mode as
\begin{equation}
\label{eq:barot}
K_{2D}=\frac12\iint\left(\left<u\right>_z^2+\left<v\right>_z^2\right) \ \textrm{d}x \ \textrm{d}y \ ,
\end{equation}
the horizontal kinetic energy associated with the fast 3D mode as
\begin{equation}
\label{eq:baroc}
K_{3D}^H=\frac12\iiint\left(u'^2+v'^2\right) \ \textrm{d}x \ \textrm{d}y \ \textrm{d}z \ ,
\end{equation}
and the vertical kinetic energy as
\begin{equation}
\label{eq:barocv}
K_{3D}^V=\frac12\iiint w^2 \ \textrm{d}x \ \textrm{d}y \ \textrm{d}z \ .
\end{equation}
Figure~\ref{fig:baro} shows the decomposition of the total kinetic energy in terms of 2D, 3D, horizontal and vertical components versus time for case $E3$ in Table~\ref{tab:one}.
Both vertical and horizontal 3D components remain small and shows no sign of growth as time increases.
The growth of the total kinetic energy is entirely due to the energy growth of the horizontal depth-independent 2D mode.

\begin{figure}
   \includegraphics[width=80mm]{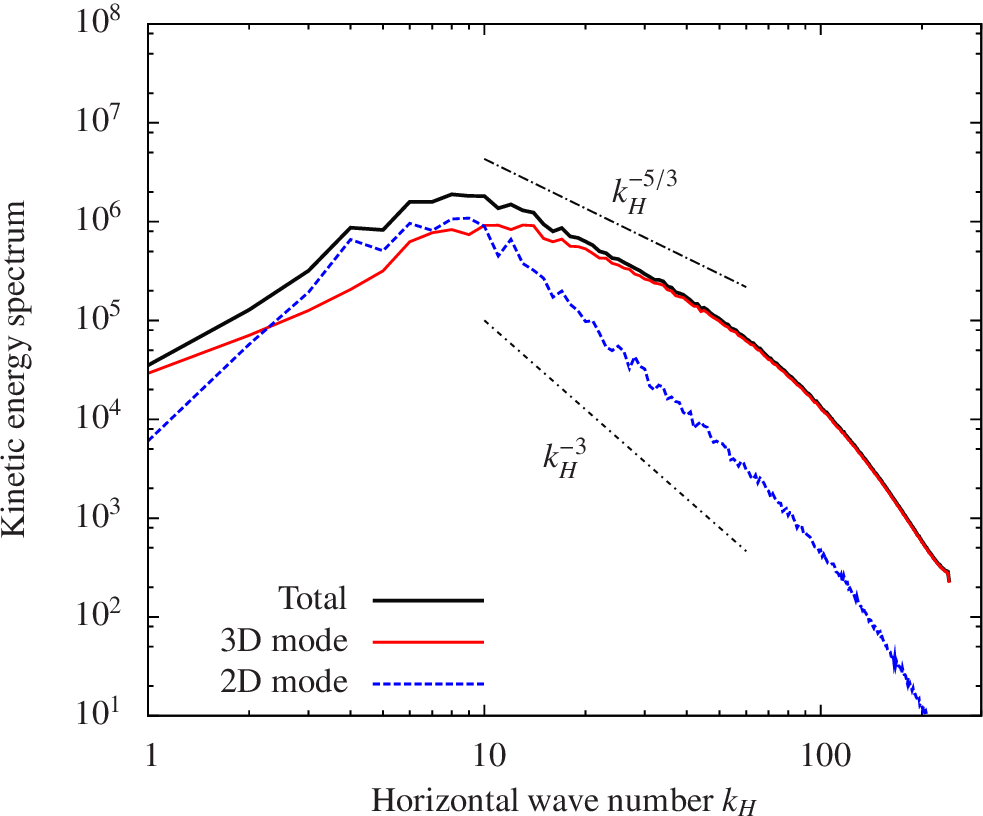}\\
   \includegraphics[width=80mm]{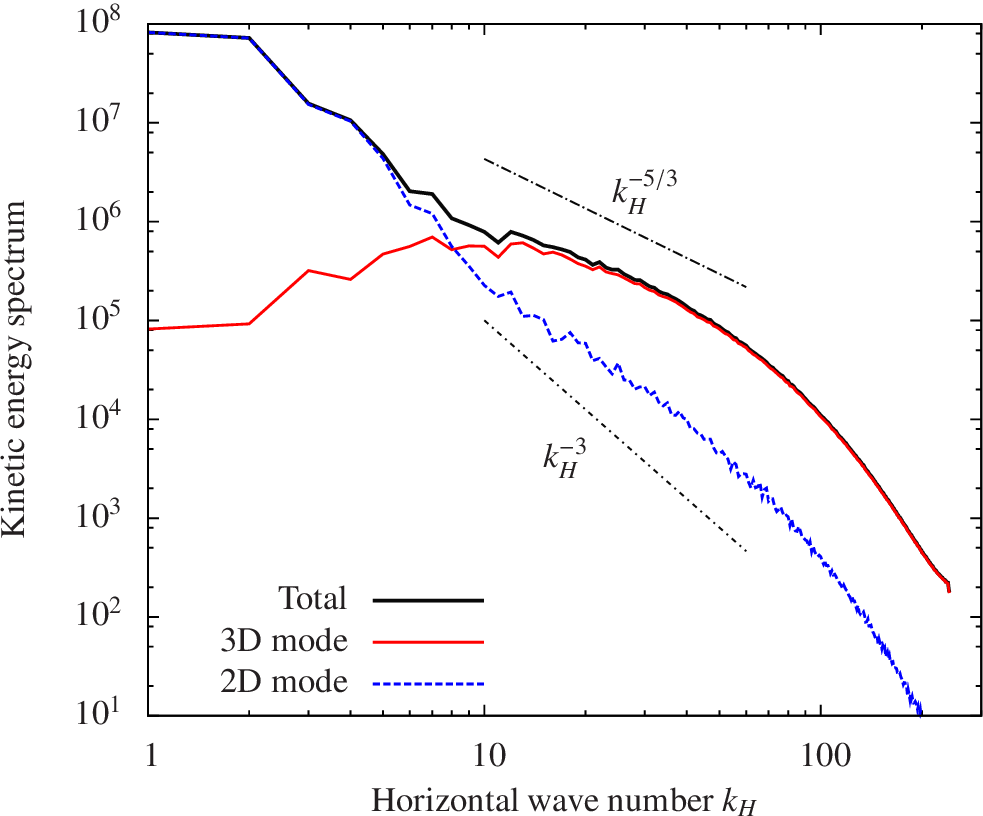}
   \caption{Horizontal kinetic energy spectrum as defined by equation \eqref{eq:spect}. Shown is the 3D mode contribution (in red), the 2D mode contribution (in blue) and the total spectrum (in black). Top: time $t=0.001$ just after the nonlinear overshoot. Bottom: time $t=0.03$ when the total kinetic energy is starting to saturate and the large-scale condensate is nearly in equilibrium.\label{fig:spect}}
\end{figure}

To gain further insight into the properties of these large-scale flows, it is useful to consider their kinetic energy spectra.
For each horizontal wave number, we define the vertically-averaged horizontal kinetic energy spectrum in the following way
\begin{equation}
\label{eq:spect}
E_K(k_H)=\frac12\sum_{z}\sum_{k_x,k_y} \hat{\bm{u}}_H(k_x,k_y,z)\cdot\hat{\bm{u}}_H^*(k_x,k_y,z)
\end{equation}
where  $\hat{\bm{u}}_H(k_x,k_y,z)$ is the 2D Fourier transform of the horizontal flow $\bm{u}_H(x,y,z)$ (with complex conjugate $\hat{\bm{u}}_H^*(k_x,k_y,z)$) and the second summation is over all $k_x$ and $k_y$ such that $k_H<\sqrt{k_x^2+k_y^2}\le k_H+1$.
Figure~\ref{fig:spect} shows the horizontal kinetic energy spectrum for both 2D (as defined by equations~\eqref{eq:vortex}-\eqref{eq:vortex2}, where the summation over $z$ in equation~\eqref{eq:spect} is not required) and 3D components (as defined by equations~\eqref{eq:upr}-\eqref{eq:wpr}) in the early stage of the nonlinear saturation and at the end of the simulation when the kinetic energy density starts to saturate.
The results correspond to simulation $E4$ in Table~\ref{tab:one}.

Figure~\ref{fig:spect} shows that initially both spectra approximately peak at scales corresponding to $k_H\approx10$.
The small scales (\textit{i.e.} $k_H>10$) are dominated by the fast 3D mode whereas both 3D and 2D modes are of comparable amplitude at larger scales.
Note that the amplitude of both the 2D and 3D modes is initially exactly zero since the fluid is at rest at $t=0$.
The 3D mode is characterized by an inertial scaling of $k_H^{-5/3}$, consistent with a 3D direct cascade of energy whereas the 2D mode displays a $k_H^{-3}$ scaling, consistent with a direct enstrophy cascade.
Similar scalings were obtained by Julien \textit{et al.}\cite{julien2012} in their geostrophic turbulence model at very low Rossby numbers, and in forced rotating and stratified turbulence in tri-periodic boxes\cite{sukhatme2008}.
At later times, the 2D vortex mode energy gradually accumulates at large horizontal scales, whereas the energy associated with the 3D mode remains qualitatively unchanged (we observe a slight increase in the 3D mode energy for $k_H<10$).
The horizontal spectrum of the vertical kinetic energy (not shown) remains qualitatively the same as the large-scale flow grows in amplitude but we note a slight decrease in the energy associated with the convective eddies at $k_H\approx10$, which is consistent with the overall reduction of the buoyancy work as previously discussed.
The flow is therefore dominated by a large horizontal scale condensate that is invariant in the $z$-direction, and virtually all the kinetic energy is concentrated in this 2D vortex mode (see Figure~\ref{fig:baro}), whereas the small-scale convective flow remains qualitatively unchanged.
We are not able to discuss the spectral behaviour at small wave numbers due to the lack of scale separation between the most unstable modes and the horizontal extent of the numerical domain. Therefore, the effect of the condensate on the spectral slope associated with the inverse cascade is not accessible here.

\subsection{Dependence on the aspect ratio and vortex merging\label{sec:ratio}}

\begin{figure}
   \hspace{-0.8cm}
   \includegraphics[width=80mm]{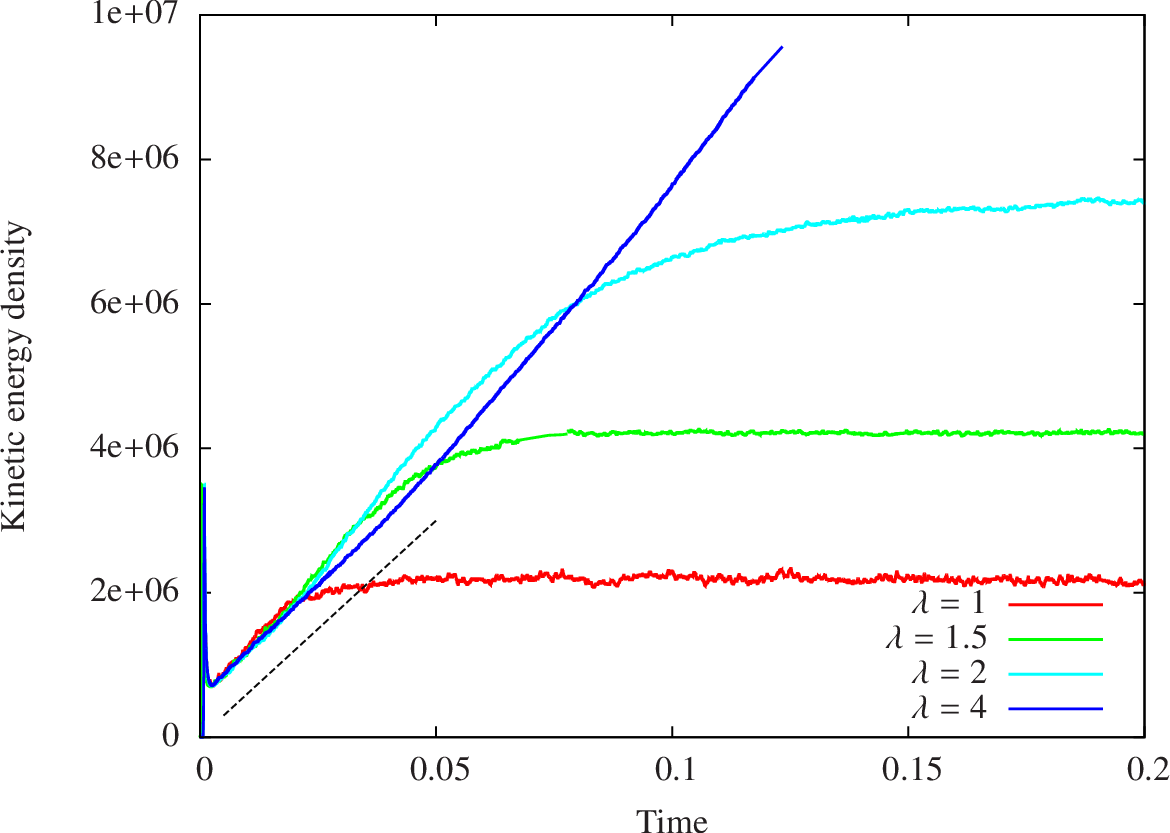}
   \caption{Time evolution of the kinetic energy density for case $E3$ varying the horizontal aspect ratio $\lambda$ of the numerical domain.\label{fig:lambda}}
\end{figure}

\begin{figure*}
   \hspace{-0.7cm}
   \includegraphics[height=43mm]{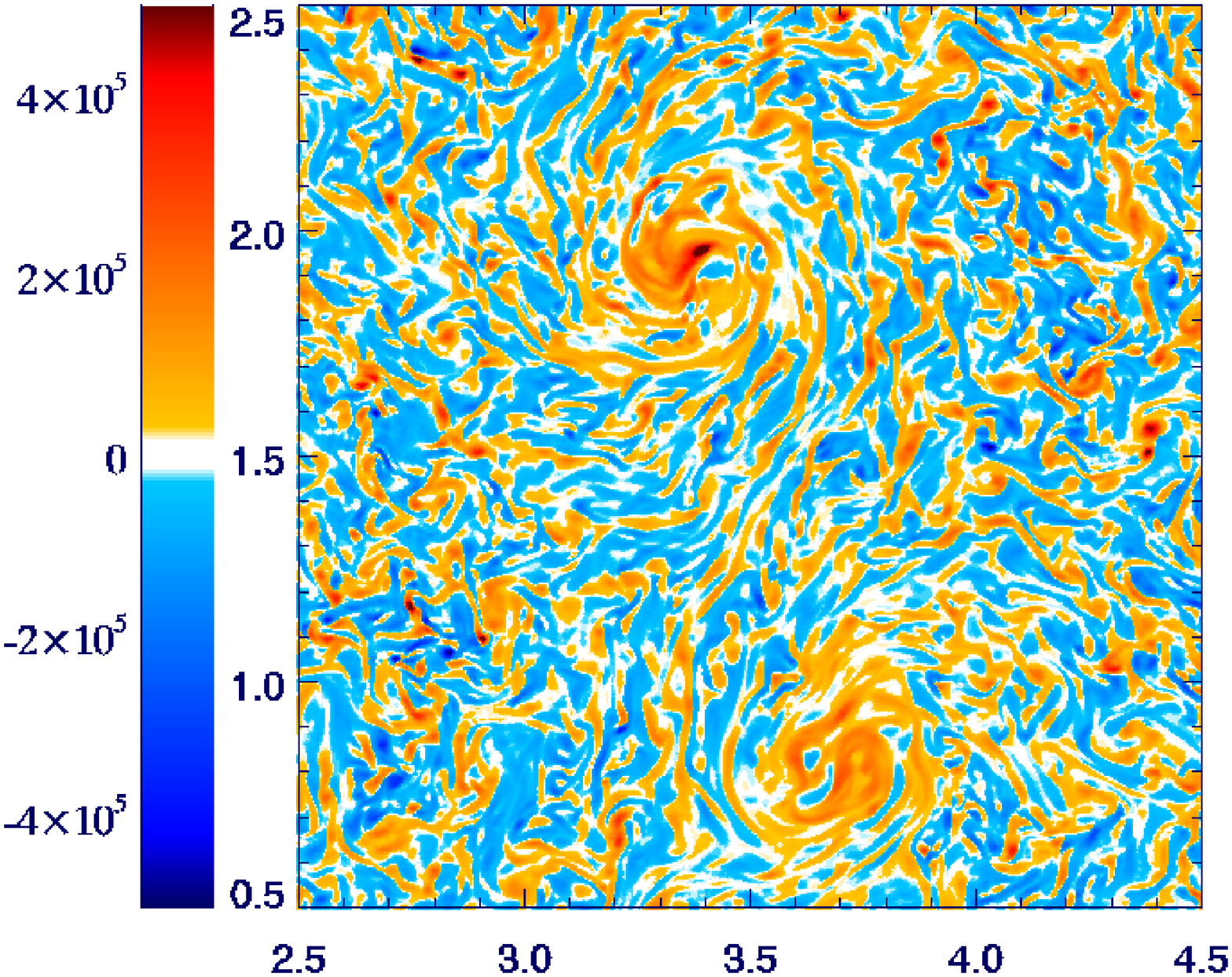}
   \includegraphics[height=43mm]{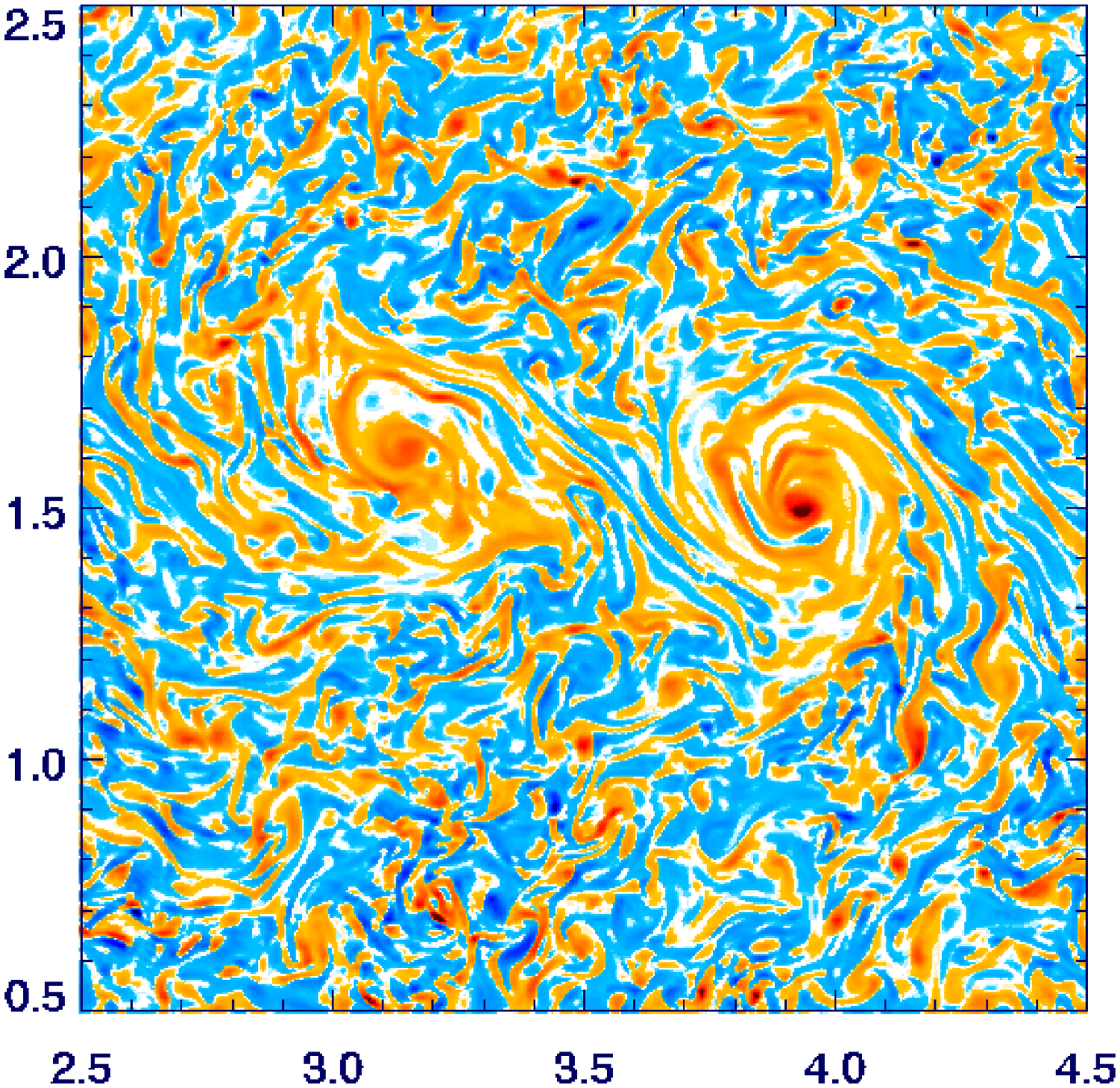}
   \includegraphics[height=43mm]{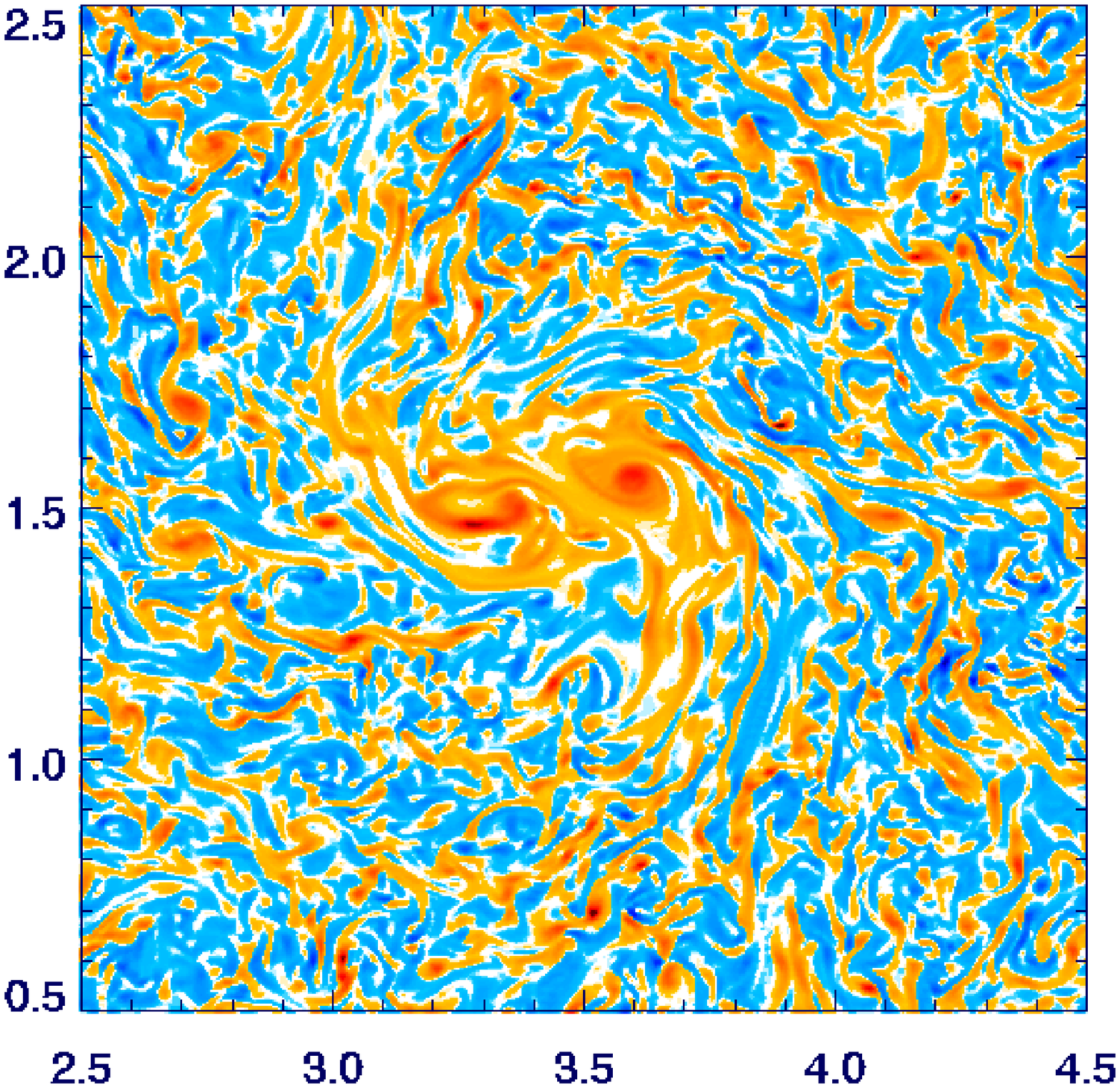}
   \includegraphics[height=43mm]{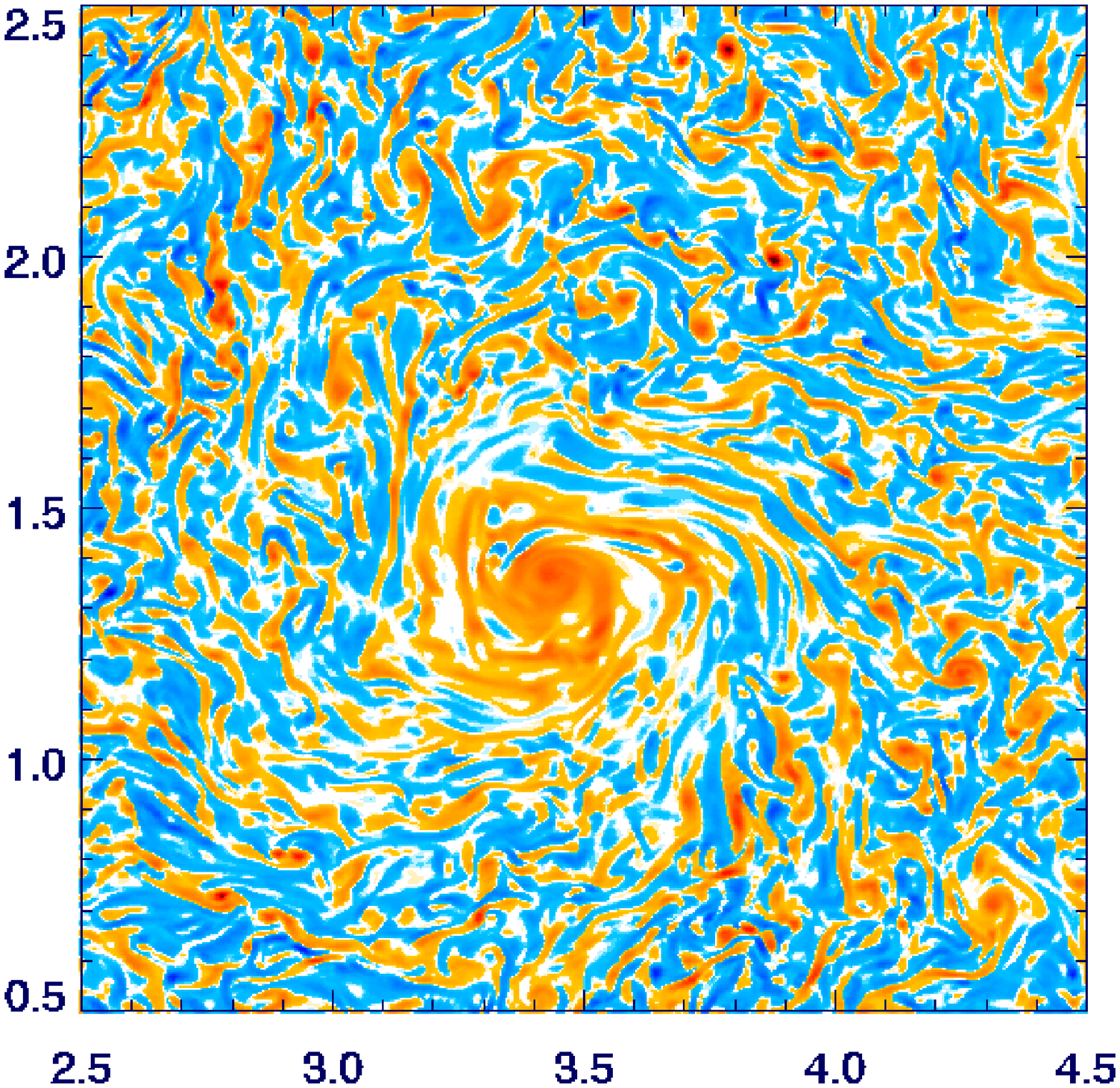}
   \caption{Vertical vorticity plotted in a horizontal plane located at $z=0.5$ for simulation $E4$. Red colours correspond to positive value of the vertical vorticity (i.e. cyclonic) whereas blue colours correspond to negative values (i.e. anti-cyclonic). Time is increasing from left to right and top to bottom, and approximately 6 turnover times separate each figure. Only $1/4\textrm{th}$ of the total horizontal numerical domain is shown.\label{fig:merg}}
\end{figure*}

A characteristic feature of the large-scale circulation is that it fills the entire numerical domain, until a so-called condensate state is reached.
In order to show this, we consider various simulations with the same set of parameters $Ta=10^{10}$ and $Ra=2\times10^8$ (case $E3$ in Table~\ref{tab:one}) but varying the horizontal aspect ratio $\lambda$ from $1$ to $4$.
The results are shown in Figure~\ref{fig:lambda}.
As $\lambda$ increases, the initial stage of the kinetic energy growth is similar for all cases, and we observe a linear growth of the total kinetic energy after the initial nonlinear overshoot.
However, the smaller the box, the quicker the circulation saturates to its equilibrium state, where the largest available horizontal scale is filled.
For the largest aspect ratio considered here, the saturation is not even reached in a run time that is numerically feasible.
The fact that the large-scale 2D circulation is a box-filling mechanism points in the direction of an upscale transfer of energy (see section \ref{sec:inv}) and similar results were reported in compressible models \cite{mantere2011}.

In low aspect ratio simulations, where there is no clear scale separation between the convective scale and the horizontal size of the box, a large-scale circulation consisting of one large cyclonic vortex gradually appears as convective cells are non-linearly interacting.
The picture is different if a true scale separation is achieved.
In this case, several cyclonic vortices of size comparable with the convective cells initially appear and slowly move relative one to another.
Eventually, two of these vortices get close together and merge in a dynamical event reminiscent of the classical vortex merging mechanism observed in 2D flows.
We show an example of such an event in Figure~\ref{fig:merg}, where a time series of the vertical vorticity in a horizontal plane located in the middle of the layer is shown.
These results correspond to the simulation $E4$ in Table~\ref{tab:one} for which $Ta=10^{10}$ and $Ra=5\times10^8$.
Approximately 6 turnover times separate each panel.

In the first panel in Figure~\ref{fig:merg}, seven cyclonic vortices are present in the whole domain (we only show $1/4\textrm{th}$ of the complete domain in order to focus on the merging event).
As time evolves, the number of vortices decreases due to merging events, as shown in the second and third panels of Figure~\ref{fig:merg}.
Eventually, only one large vortex remains at the end of the computation.
This vortex merging mechanism is only observed in large aspect ratio simulations and is probably not enough in itself to explain the origin of the large-scale circulation since low aspect ratios simulations can still sustain a large-scale circulation although no vortex merging events are observed in this case.
In addition, the first initial vortices are spontaneously formed in the layer without any merging events to generate them.

\subsection{Dependence on the parameters}

\begin{figure}
   \hspace{-0.4cm}
   \includegraphics[width=85mm]{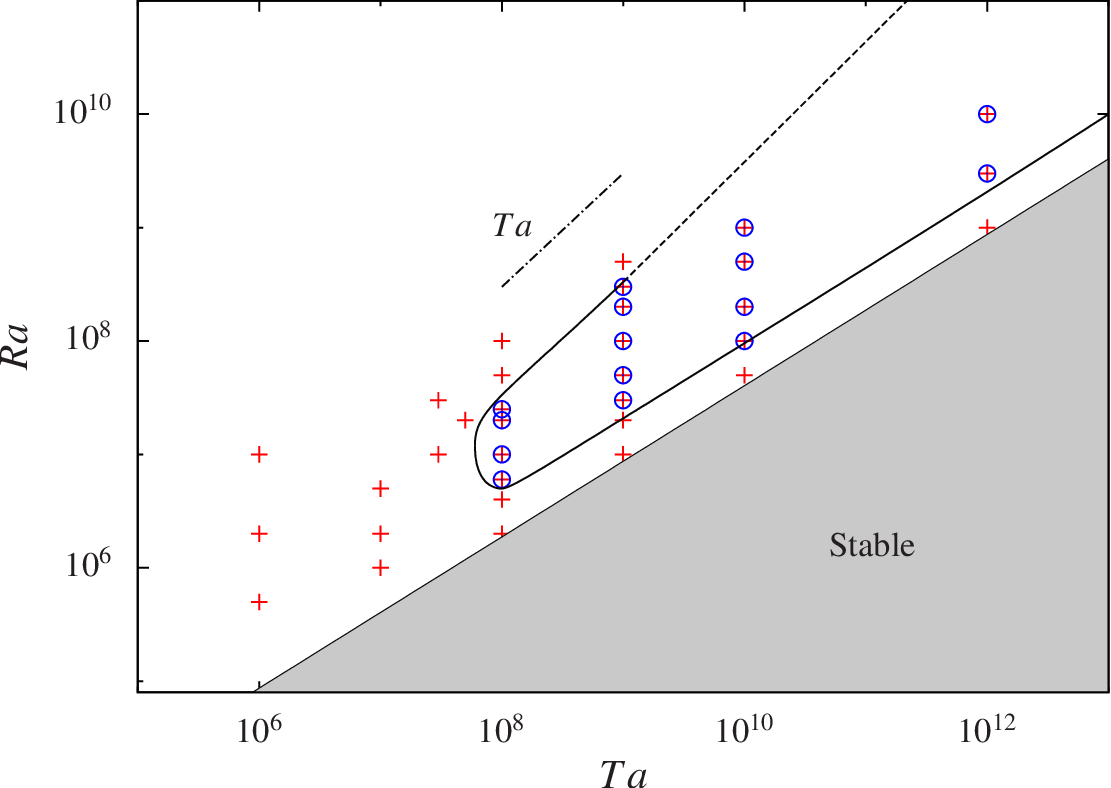}
   \caption{Summary of the Boussinesq simulations considered in this paper in the plane $(Ta,Ra)$. The grey area corresponds to the stable regime. The critical Rayleigh number is approximately given by $Ra_c\approx Ta^{2/3}$. Blue circles correspond to simulations where a large-scale circulation is observed. The domain of existence in parameter space of the large-scale circulation is delimited by the thick black line. The dotted line corresponds to a suggested boundary in the large $Ra, Ta$ regime which is not accessible numerically. The dot-dash line is showing the scaling $Ra\approx Ta$ and corresponds to a constant value of the Rossby number around unity as defined by equation~\eqref{eq:ro}.\label{fig:param}}
\end{figure}

For all of the simulations discussed in this paper, we assume that there is a large-scale circulation as soon as we observe an increase in the kinetic energy associated with the 2D mode, as seen in Figure~\ref{fig:baro}.
Note that for some simulations (case $E2$ for example), the growth rate of the kinetic energy of the 2D mode is very small when compared to the turnover time of the small-scale turbulent flow and so a very long time integration is required to unambiguously identify the 2D mode.
Simulations displaying such a large-scale horizontal flow are shown by a blue circle in Figure~\ref{fig:param}.
Clearly, only simulations sufficiently far from the onset of convection, \textit{i.e.} in a turbulent state, can sustain a large-scale flow.
This is further confirmed by looking at the local Reynolds number of the flow, as defined by equation~\eqref{eq:re} in Table~\ref{tab:one}.
Only the simulations for which $Re>20$ are able to sustain a large-scale circulation.
In addition, only simulations with a low enough Rossby number, typically $Ro_L<1$, show a growth in the kinetic energy associated with the 2D mode.

The two conditions described above, together with the fact that the aspect ratio must be large enough to accommodate for the large-scale flow (see section \ref{sec:ratio}), are difficult to satisfy numerically, which could explain why this behaviour was only observed relatively recently in numerical models\cite{chan2007}.
It is relatively easy to obtain a turbulent flow in a classical Rayleigh-B\'enard set-up by increasing the Rayleigh number to highly supercritical values.
However, this tends to decrease the effect of rotation (or equivalently increase the Rossby number) so that no large-scale circulation can be observed.
Maintaining a regime with both large Reynolds number and low Rossby number requires very large rotation rates, so that even when the Rayleigh number is supercritical and the flow turbulent, the Rossby number remains small compared to unity.
These results are qualitatively consistent with previous parametric studies performed in the fully-compressible regime \cite{mantere2011}.

For each value of the Taylor number larger than $Ta=10^8$, we found a critical Rayleigh number above which a large-scale vortex mode is sustained.
This threshold Rayleigh number approximately corresponds to five times the critical value $Ra_c$.
As the Rayleigh number is further increased while the Taylor number is fixed, we report a second transition above which the vortex mode is not sustained any more.
We managed to reach this regime numerically for two Taylor numbers, $Ta=10^8$ and $Ta=10^9$, where we found growth of the vortex mode at moderate Rayleigh numbers until the large-scale flow is no longer sustained at very large Rayleigh numbers (cases $C5$, $C6$ and $D6$ in Table~\ref{tab:one}).
The dotted line in Figure~\ref{fig:param} corresponds to this transition and our data suggest a critical Rayleigh number for the disappearance of the vortex mode scaling as $Ta$.
Note that this scaling corresponds to a constant value of the Rossby number using the definition given by equation~\eqref{eq:ro}.
This confirms that the large-scale flow is no longer sustained as soon as the Rossby number exceeds a critical value close to unity.

The fact that we only observe a large-scale circulation for a given range of parameters indicates that there is a transition between two different states of the flow as both Taylor and Rayleigh numbers are varied.
This transition is linked to the anisotropy of the flow, which strongly depends on the input parameters as we will discuss below.
Anisotropy in rotating Boussinesq convection has been studied by Kunnen \textit{et al.}\cite{kunnen2008,kunnen2010}.
Following their approach, we use the formalism introduced by Lumley\cite{lumley1977,choi2001} in order to characterise the anisotropy of a turbulent flow in terms of the properties of the Reynolds stress tensor.
We define here the Reynolds stress tensor as $R_{ij}=\left<u'_iu'_j\right>$ where the brackets now denote a temporal and horizontal average.
Here, $u'_i$ is the velocity component of the 3D mode only in the $i$th direction, as defined by equations \eqref{eq:upr}-\eqref{eq:wpr}.
The Reynolds stress tensor of the total velocity field is indeed strongly time-dependent (due to the slow evolution of the large-scale circulation) whereas the tensor associated with the 3D mode only is quasi-stationary, even in the case where a large-scale circulation is sustained (see, for example, the time evolution of the kinetic energy associated with the 3D mode in Figure~\ref{fig:baro}).

\begin{figure}
   \hspace{-0.25cm}
   \includegraphics[height=60mm]{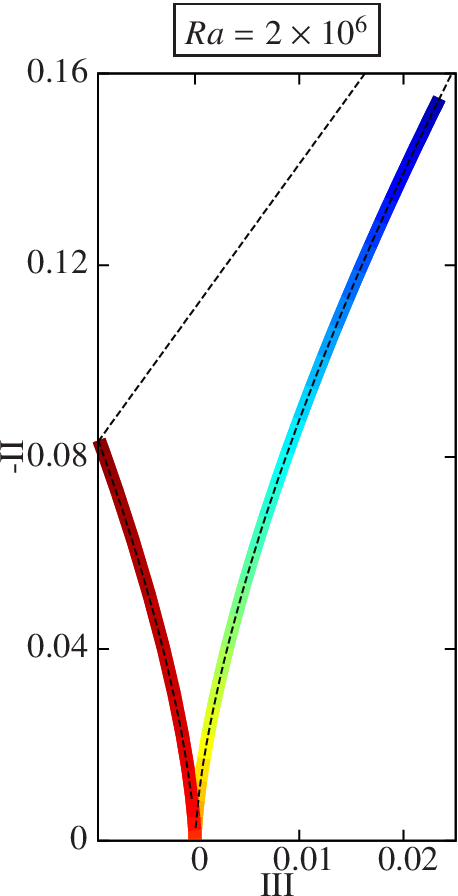}
   \includegraphics[height=60mm]{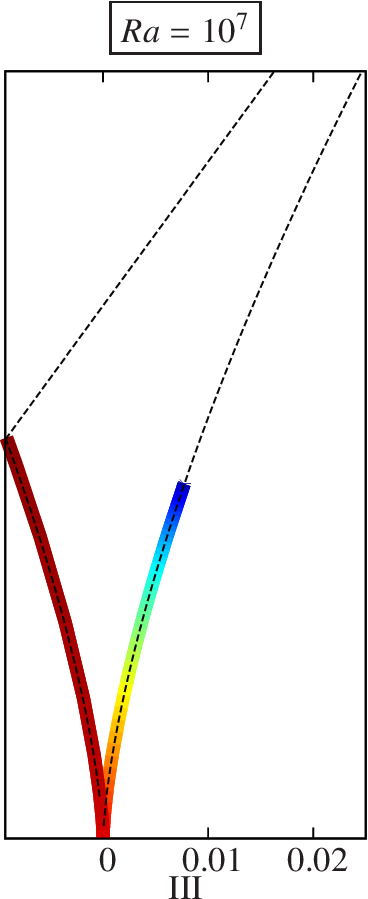}
   \includegraphics[height=60mm]{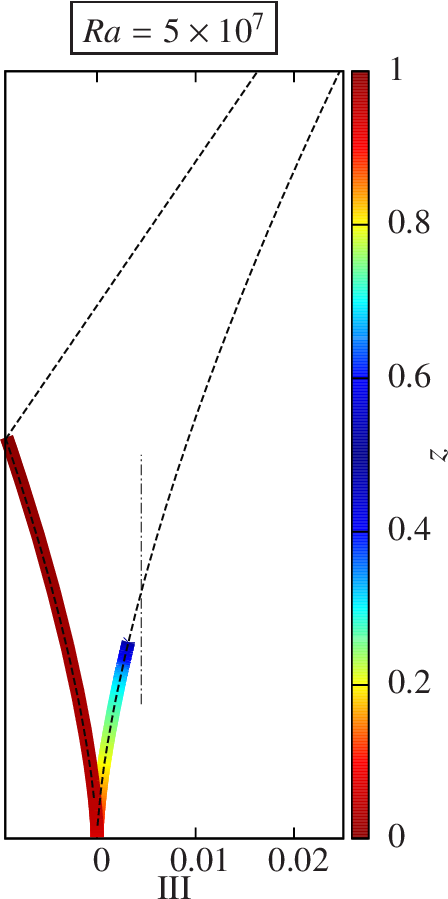}
   \caption{Invariants of the deviatoric part of the Reynolds stress tensor plotted in a so-called Lumley map for cases $C1$, $C3$ and $C5$. Red corresponds to the region close to the boundaries whereas blue corresponds to the middle of the layer. From left to right, the Rayleigh number is increasing. The vertical dash-dotted line in the right panel correspond to the approximate critical value of the invariant $I\!I\!I$ below which the flow cannot sustain a 2D mode.\label{fig:bij}}
\end{figure}

The anisotropy of the flow can be characterised by the deviatoric part of $R_{ij}$, defined as $b_{ij}=R_{ij}/R_{kk}-\delta_{ij}/3$ where $\delta_{ij}$ is the Kroenecker tensor and summation is implied over repeated indices.
The tensor $b_{ij}$ is traceless and symmetric, and has been used to describe the anisotropy in physical space of many turbulent flows.
Note that the tensor $b_{ij}$ is also related to a more complete two-points homogeneous description of the anisotropy in spectral space (see for example Cambon \textit{et al.} \cite{cambon1997}).
The three invariants of the tensor $b_{ij}$ are
\begin{equation}
I=b_{ii}=0 \ , \quad I\!I=\frac{-b_{ij}b_{ji}}{2} \quad \textrm{and} \quad I\!I\!I=\textrm{Det}\left(b_{ij}\right) \ .
\end{equation}
The second and third invariants provide a simple way to describe the one-point anisotropy of the flow in the plane $(I\!I\!I, -I\!I)$.
Several limiting cases are of interest: when $I\!I\!I=I\!I=0$, the flow is effectively isotropic in the sense that the Reynolds stress tensor is isotropic.
There is a limiting curve on the left of the map (defined by $I\!I\!I=-2\left(-I\!I/3\right)^{3/2}$, see \cite{choi2001} for more details) which corresponds to what has been called ``pancake shaped'' 2D turbulence and corresponds to the case where two eigenvalues of the deviatoric tensor $b_{ij}$ are dominant.
The second limiting curve on the right part of the map corresponds to $I\!I\!I=2\left(-I\!I/3\right)^{3/2}$ and is often referred as to ``cigar-shaped'' 2D turbulence since one eigenvalue is dominant.

Figure~\ref{fig:bij} shows the two invariants in a Lumley map for different Rayleigh numbers and a fixed Taylor number of $Ta=10^8$ (cases $C1$, $C3$ and $C5$ in Table~\ref{tab:one}).
As expected, in all cases, the turbulence is 2D and dominated by the two horizontal components very close to the boundaries $z=0$ and $z=1$ (red colour, limiting curve on the left in Figure~\ref{fig:bij}).
This state of ``pancake'' structures, forced by the presence of the solid boundaries, rapidly shifts towards a more isotropic state as we move away from the boundaries.
At very low Rayleigh numbers (left panel in Figure~\ref{fig:bij}), the anisotropy of the flow in the middle of the layer is characterized by 2D ``cigar'' shaped structures where the vertical velocity is dominating.
As the Rayleigh number is increased from case $C1$ to case $C5$, this strong anisotropy is reduced, so that at sufficiently large Rayleigh number, the flow in the middle of the fluid layer tends towards a quasi-isotropic 3D state ($I\!I\!I \approx I\!I \approx 0$).

\begin{figure}
   \hspace{-0.25cm}
   \includegraphics[height=21mm]{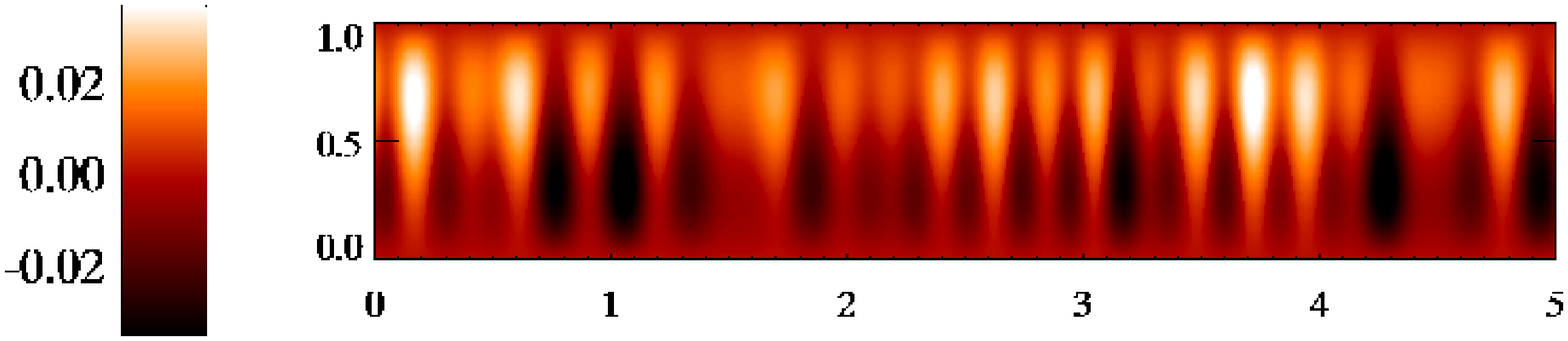}\\
   \includegraphics[height=21mm]{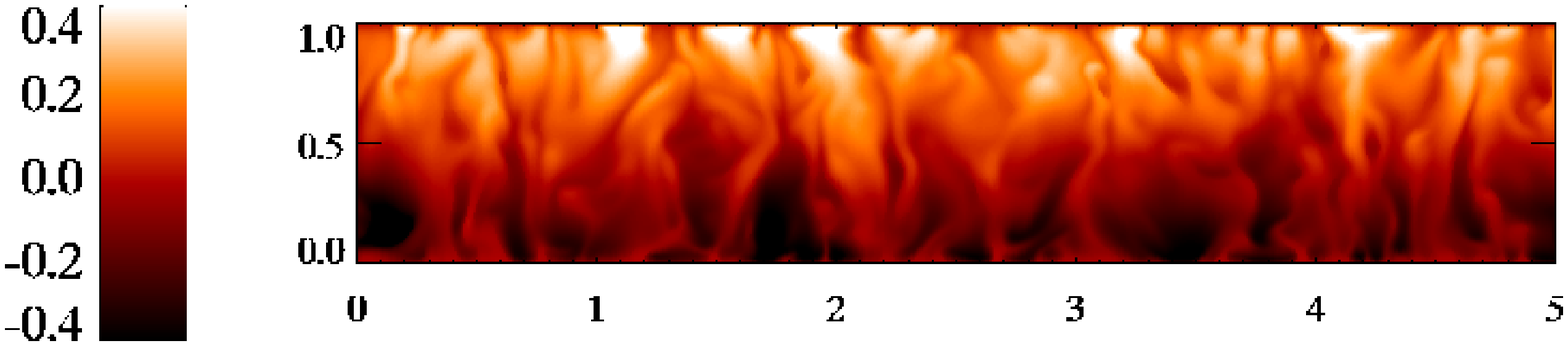}\\
   \includegraphics[height=21mm]{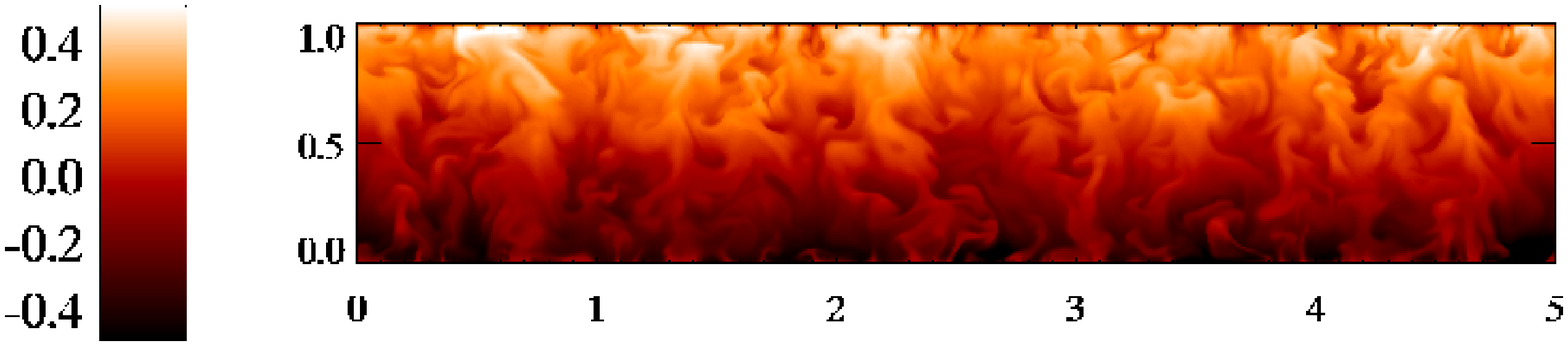}
   \caption{Vertical slice in a $(x,z)$ plane where temperature fluctuations are shown. Bright colours correspond to positive temperature fluctuations whereas dark colours correspond to negative fluctuations. The Taylor number is $Ta=10^8$ and the Rayleigh number is $2\times10^6$, $10^7$ and $5\times10^7$ from top to bottom (simulations $C1$, $C3$ and $C5$ in Table~\ref{tab:one}). Only the case in the middle is able to sustain a large-scale circulation.\label{fig:side}}
\end{figure}

For $Ra=2\times10^6$, the flow in the bulk of the layer is strongly anisotropic, but the Reynolds number is very low ($Re=0.67$, see case $C1$ in Table~\ref{tab:one}).
The Rayleigh number is very close to its critical value ($Ra/Ra_c=1.1$) so that only a very limited range of horizontal wave numbers are unstable.
In that case, convective columns barely interact, the flow is quasi-stationary and no large-scale flow is observed (see Figure~\ref{fig:param}).
For $Ra=10^7$, the bulk flow is still anisotropic (see Figures~\ref{fig:bij} and \ref{fig:side}) but the Reynolds number is now large enough ($Re=43.8$, see case $C3$ in Table~\ref{tab:one}) to allow for non-linear interaction between the larger range of unstable convective cells, leading to the formation of a large-scale circulation.
Finally, for $Ra=5\times10^7$, the anisotropy of the flow is strongly reduced (apart from the region very close to the boundaries).
Thermal boundary layers are destabilized and coherent localised thermal plumes are emitted from them, resulting in a turbulent flow in the bulk of the layer in a quasi-isotropic 3D state.
The energy associated with the 2D mode remains small compared to the other components in this case.
Note that for cases with $Ta=10^8$ and $Ta=10^9$, we found a maximum value of the Rayleigh number above which no large-scale flow is sustained due to the reisotropization of the flow (see Figure~\ref{fig:param} and Table~\ref{tab:one}).
In both cases, the transition occurs for a critical value of the third invariant $I\!I\!I$ around $0.04$, as shown by a vertical dashed line in the right part of Figure~\ref{fig:bij}.

The analysis above is confirmed by visually looking at the flow properties as the Rayleigh number is increased for a fixed Taylor number.
Figure~\ref{fig:side} shows temperature fluctuations in a vertical $(x,z)$ plane for the same three Rayleigh numbers $Ra=2\times10^6$, $10^7$ and $5\times10^7$ and $Ta=10^8$.
At the lowest Rayleigh number, laminar quasi-steady columnar convective cells are found.
As the Rayleigh number is increased, localised thermal plumes start to appear that eventually lead to a fully-turbulent regime.
Only for the intermediate Rayleigh number, where the turbulent in intense enough but the flow is still quasi-2D, is the large-scale flow sustained.
Note that in the convective cell regime, the relative helicity of the flow is nearly maximum whereas it is close to zero in the thermal plumes regime.

The transition discussed above in terms of anisotropy corresponds to the transition empirically measured in terms of heat flux by Schmitz \textit{et al.}\cite{schmitz2010}.
They found a transition around $Re_L \sigma Ta^{-1/4}\approx 10$, where $Re_L$ is the Reynolds number based on the root-mean square velocity and the depth of the layer and $\sigma$ is the Prandtl number.
For $Ta=10^8$, the transition occurs between cases $C4$ and $C5$, where $Re_L \sigma Ta^{-1/4}$ is varying from $6.2$ for case $C4$ to $10.7$ for case $C5$.
For $Ta=10^9$, the transition occurs between cases $D5$, for which $Re_L \sigma Ta^{-1/4}=11.4$, and case $D6$, for which $Re_L Pr Ta^{-1/4}=16.8$.
This indicates that the transition discussed by Schmitz \textit{et al.}\cite{schmitz2010} in terms of heat flux is the same as the transition discussed here in terms of large-scale circulation.
Note however that these authors did not mention the existence of a large-scale flow in their simulations.

\subsection{Inverse energy transfer\label{sec:inv}}

Using a reduced set of equations\cite{julien1996}, it has been suggested that the origin of the large-scale circulation is due to a non-local inverse cascade of energy \cite{julien2012}.
These authors decomposed the flow into barotropic (\textit{i.e.} 2D) and baroclinic (\textit{i.e.} 3D) components and described the self-interaction of the barotropic component and the cross-interaction between the two components.
Here, we provide another proof for non-local inverse cascade of kinetic energy using the full Boussinesq equations, valid for any values of the Rossby number, and a different shell-to-shell transfer analysis.
We use the formalism described in several papers\cite{alexakis2005,mininni2005} and later used to study a variety of homogeneous turbulent flows, from rotating \cite{mininni2009} to magnetohydrodynamic turbulence \cite{alexakis2007}.

We first compute the 2D horizontal Fourier transform of the velocity field $\hat{\bm{u}}(k_x, k_y, z)$.
We then define the filtered velocity field $\bm{u}_K(\bm{x})$ as
\begin{equation}
\bm{u}_K(x,y,z)=\sum_{K<k_H \le K+1}\hat{\bm{u}}(k_x,k_y,z)e^{ik_xx}e^{ik_yy} \ ,
\end{equation}
where $k_H=\sqrt{k_x^2+k_y^2}$ is the horizontal wave number.
This velocity field in physical space corresponds to the sum of all modes with a horizontal wave vector lying in the cylindrical shell $K<k_H\le K+1$.
Note that by summing all filtered velocity fields from $K=0$ to $K=k_{\textrm{max}}$, we recover the initial total velocity field $\bm{u}$.
The evolution of the kinetic energy in shell $K$, defined by
\begin{equation}
E(K)=\frac12\int_V|\bm{u}_K|^2\textrm{d}V
\end{equation}
is given by
\begin{equation}
\frac{\partial E(K)}{\partial t}=\sum_Q \mathcal{T}(Q,K)-\mathcal{D}(K)+\mathcal{F}(K)
\end{equation}
where the function $\mathcal{T}(Q,K)$ corresponds to the energy transfer from shell $Q$ to shell $K$.
By summing this function over all possible shells $Q$, we get the total energy transfer in and out of shell $K$.
$\mathcal{T}(Q,K)$ is defined by
\begin{equation}
\mathcal{T}(Q,K)=-\int_V \bm{u}_K\cdot\Big(\bm{u}\cdot\nabla\bm{u}_Q\Big)\textrm{d}V \ ,
\end{equation}
and it satisfies the symmetry property $\mathcal{T}(Q,K)=-\mathcal{T}(K,Q)$.
If $\mathcal{T}(Q,K)>0$, then a positive amount of kinetic energy is extracted from shell $Q$ and given to shell $K$.
The kinetic energy dissipation rate is defined by
\begin{equation}
\mathcal{D}(K)=\sigma\int_V |\nabla\bm{u}_K|^2\textrm{d}V \ ,
\end{equation}
and the buoyancy work is defined by
\begin{equation}
\mathcal{F}(K)=\sigma Ra \int_V\theta\bm{u}_K\cdot\bm{e}_z\textrm{d}V \ .
\end{equation}
The buoyancy work $\mathcal{F}(K)$ is always negligible at small wave numbers, so that the kinetic energy deposited there must come from another mechanism.

\begin{figure}
   \includegraphics[width=80mm]{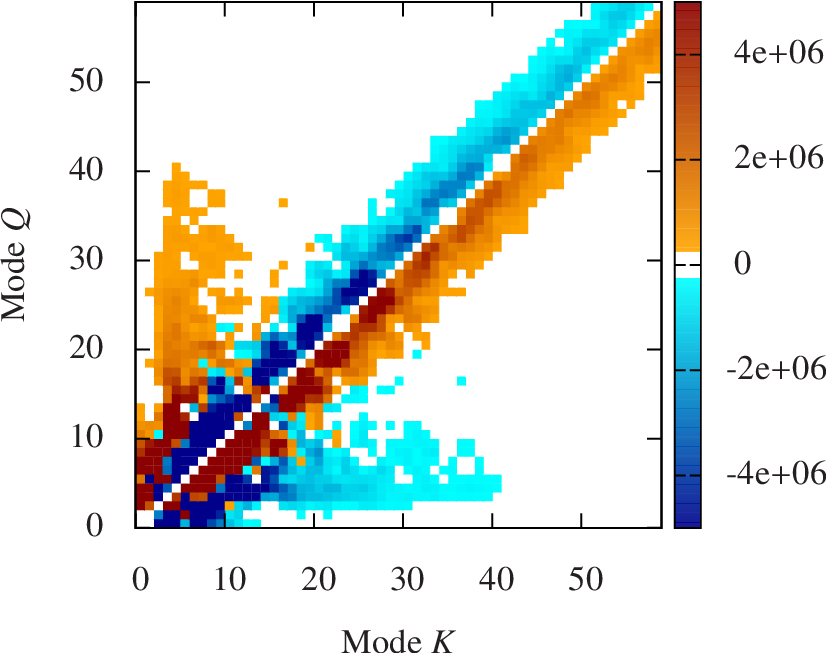} \\
   \includegraphics[width=80mm]{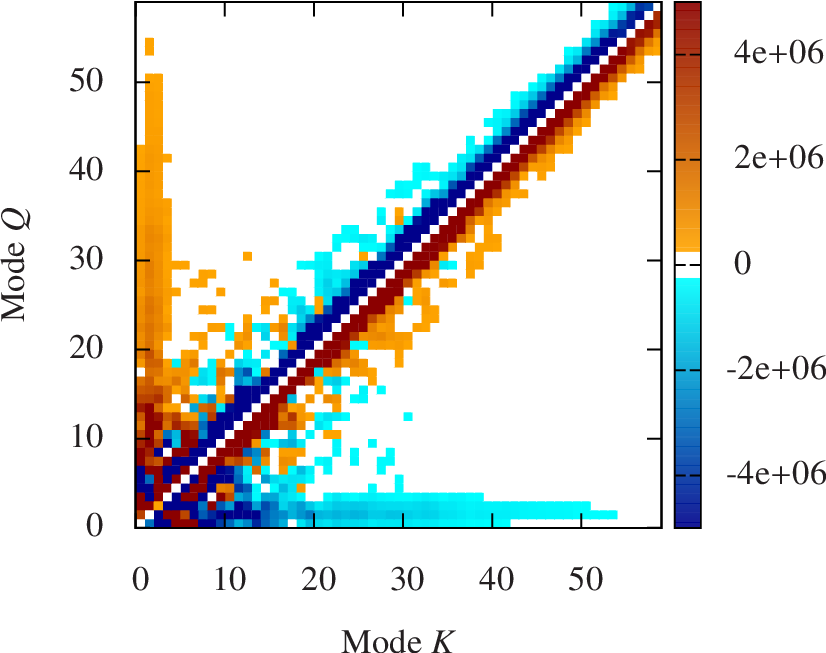}
   \caption{Kinetic energy transfer function $\mathcal{T}(Q,K)$ for shells $0<Q,K<59$ and case $E4$ in Table~\ref{tab:one}. The function is positive when energy is taken from shell $Q$ and negative when energy is received. The top figure corresponds to $t\approx0.01$ (just after the initial nonlinear saturation) whereas the bottom figure corresponds to $t\approx0.04$ (when the large-scale flow saturates at the box size), see Figure~\ref{fig:kin}.\label{fig:tr}}
\end{figure}

We calculate the vertically averaged transfer function for simulation $E4$.
Results are shown in Figure~\ref{fig:tr} at two different times, when the vortex mode starts growing (\textit{i.e.} at $t\approx0.01$, see Figure~\ref{fig:kin}) and when it saturates at the box size (\textit{i.e.} at $t\approx0.04$).
For this particular simulation, the convective scale before the appearance of a large-scale circulation approximately corresponds to $k_f=10$.
For shells $Q\approx K$ (\textit{i.e.} close to the diagonal) and at scales smaller than the convective scale, the transfer is mostly local (\textit{i.e.} concentrated close to the diagonal).
Energy is extracted from shell $K$ slightly smaller than shell $Q$ and given to slightly larger scales $K$.
This is a characteristic of a direct local energy transfer between shells.
As time evolves, this direct energy transfer becomes more and more localised around the diagonal, showing that the direct cascade of energy at small scales is mediated by interactions with the largest scale in the system $k_H\approx1$.
This is consistent with the cascade mechanism observed in forced rotating turbulence\cite{mininni2009}.
In addition, we also observe a positive transfer in the region $K<10$ and $Q>10$.
Thus the small convective scales are loosing energy in favour of large-scale scale flow.
This is typical of an inverse non-local energy transfer from the small convective cells to the large scales.
The transfer is non-local since there can be a significant difference between the wavenumber of the shell giving energy and the wavenumber of the shell receiving energy. 
Note that these non-local transfers are not observed if we compute the energy transfer function associated with the wave mode only (not shown).
In this case, only a direct local cascade from the most unstable scale down to the dissipative scale is observed.
The transfer function shown in Figure~\ref{fig:tr} is very similar to the results discussed by Mininni \textit{et al.}\cite{mininni2009} in the context of rotating turbulence and using the same formalism.
An alternative analysis of the energy transfers at low Rossby numbers can be found in Rubio \textit{et al.}\cite{rubio2014} and filtered simulations have been recently discussed by Guervilly \textit{et al.}\cite{guervilly2014}.

The existence of an inverse cascade of energy in the context of constrained 3D flows is a long-standing issue \cite{biferale2012}.
While the particular case of 2D flows has been thoroughly studied, there is far less examples of incoherent inverse cascade in fully-developed three dimensional flows.
One way to gradually transit from 3D to 2D behaviours is to spatially confine the flow in one direction until an inverse cascade eventually appears.
This approach has been taken by several authors\cite{smith1996,celani2010}, and the results are very similar to our set of simulations with constant Taylor number but increasing Rayleigh number.
We indeed obtain a regime where a linear growth of the kinetic energy is observed (see Figure~\ref{fig:lambda}) until the flow saturates at the box scale.
However, the transition from 3D to 2D dynamics naturally appears in our system due to the coupled action of rotation, buoyancy and horizontal boundaries, as opposed to a prescribed spatial confinement only.
Note further that Smith \textit{et al.}\cite{smith1996} considered the simultaneous effects of both rotation and reduction of one of the spatial dimensions, which is very similar to our case.

As discussed earlier, the existence of an inverse cascade is, in our case, related to the value of the Rossby and Reynolds numbers.
If the Rossby number is too large, the flow is dominated by the buoyancy and is mostly 3D apart from the thin boundary layers.
This is the case for flows with $Ta<10^8$ at any Rayleigh numbers.
The anisotropy is moderate and the 3D flow can only cascade energy from the injection scale down to the viscous dissipation scale.
If the Rossby number is small enough, the flow is dominated by the Coriolis force, which enhances the horizontal mixing in the layer.
The bulk of the flow is quasi-2D, so that an inverse cascade can eventually develop in addition to the direct cascade, sustaining a large-scale depth-invariant horizontal flow.
Although this mechanism is robust, it only exists in a rather limited region of the parameter space (see Figure~\ref{fig:param}), since large enough Rayleigh numbers will eventually lead to a return to isotropy and associated downscale cascade.

\subsection{Cyclones, anti-cyclones and symmetry breaking\label{sec:cycl}}

In most of the results discussed so far, there is a clear symmetry breaking in the large-scale circulation.
We observe a dominance of cyclonic structures in favour of anti-cyclonic structures (see for example Figure~\ref{fig:merg}).
Cyclonic vortices are rotating in the same direction than the background rotation and correspond here to positive values of the vertical vorticity.
Anti-cyclonic structures are rotating in the opposite direction and correspond to negative vertical vorticity.
In the case of no-slip boundary conditions, the prevalence of cyclonic thermal plumes can be explained by the Ekman pumping at the boundaries \cite{julien1996}.
In our stress-free case however, the corresponding pumping is much weaker so that another explanation must be provided for the existence of preferentially cyclonic plumes.

To try to understand the symmetry breaking in our simulations it is instructive to look at the probability density function (PDF) of the vertical vorticity in various simulations.
As already observed by several authors\cite{julien1996}, the PDF of vertical vorticity is strongly skewed close to the boundaries, where strong positive values are more likely.
A positive skewness for the vertical vorticity implies a dominance of large positive vorticity events.
This is a property of rotating convection at moderate Rossby number even in the absence of large-scale circulation.
As the Rossby number is decreasing, the symmetry between cyclonic and anti-cyclonic structures is restored.
In the limit of vanishing Rossby number, asymptotic models predict a symmetric state between both cyclonic and anti-cyclonic structures\cite{julien2012}.

\begin{figure}
   \hspace{-0.8cm}
   \includegraphics[width=80mm]{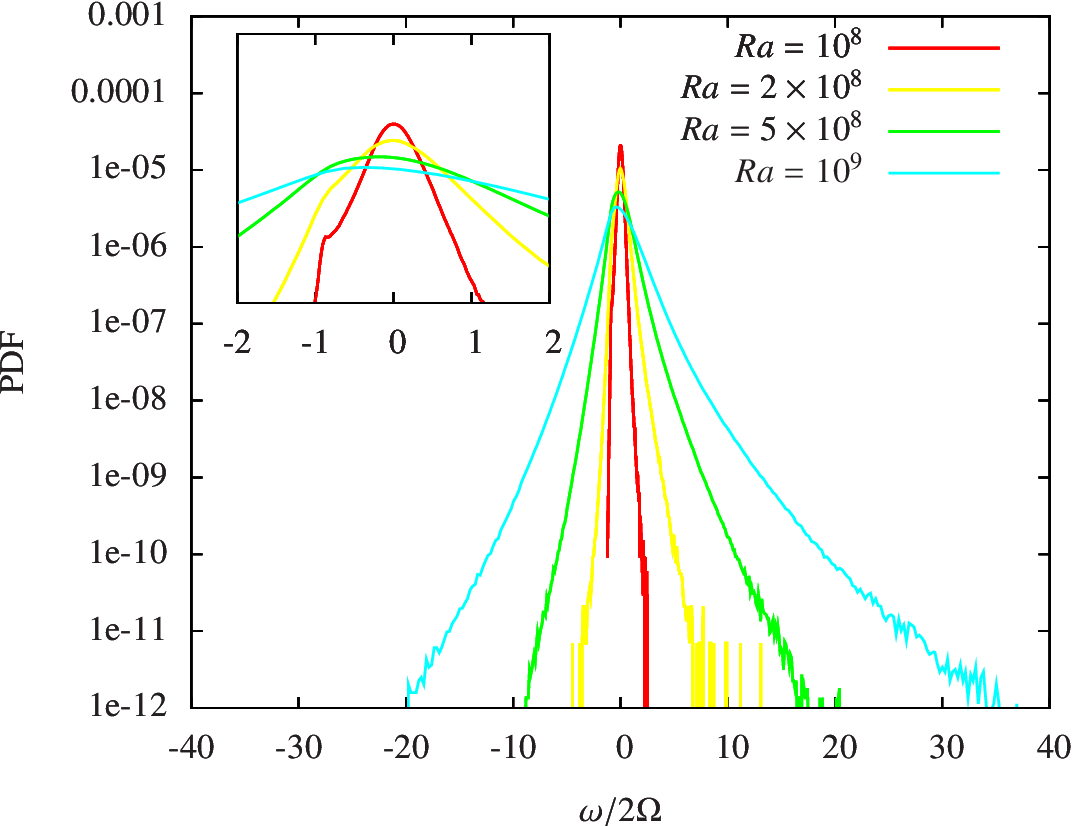}\\
   \includegraphics[width=80mm]{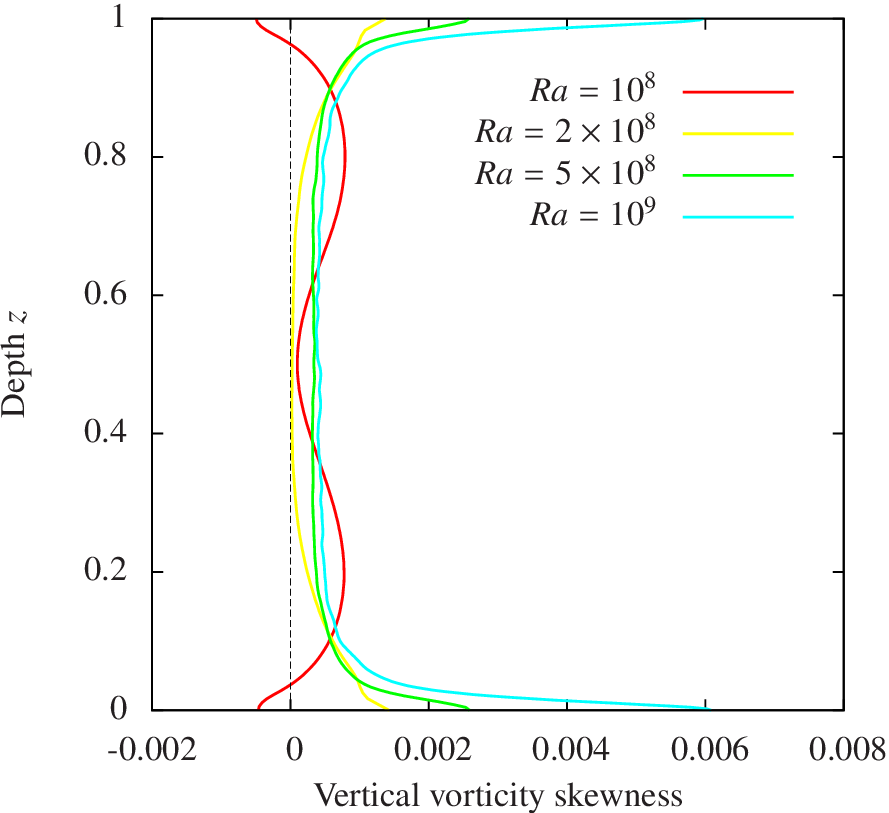}
   \caption{Top: Probability density function of the vertical vorticity in the region close to the boundaries. The vertical vorticity is normalized by the background value $2\Omega$. The insert plot is a close up of the region $-2<\omega/2\Omega<2$, where a transition is observed at the value $\omega/2\Omega=-1$. Bottom: Vertical vorticity skewness as a function of depth. The results correspond to cases $E2$ to $E5$ in Table~\ref{tab:one}.\label{fig:pdfw}}
\end{figure}

Figure~\ref{fig:pdfw} shows the PDF of the vertical vorticity close to the boundaries.
The Taylor number is fixed to $Ta=10^{10}$ and we consider different values of the Rayleigh number.
In all cases, the PDF is strongly skewed in favour of large positive values of the vertical vorticity.
As the Rayleigh number is decreased (and the Rossby number decreased), the PDF becomes more and more symmetric.
The skewness of the vertical vorticity, defined by
\begin{equation}
S_{\omega}=\frac{\left<\omega_z^3\right>}{\left<\omega_z^2\right>^{3/2}}
\end{equation}
is also plotted in the bottom panel of Figure~\ref{fig:pdfw} as a function of depth.
As already observed in previous studies\cite{julien1996}, the skewness of the vertical vorticity is positive close to the boundaries and remains very small in the bulk of the layer.
Again, we emphasize that these observations do not depend on the existence, or lack of, a large-scale circulation.
The vorticity statistics are nearly stationary even during the growth phase of the 2D mode.
As the 2D mode saturates at the box scale, we observe a slight decrease of the vertical vorticity skewness in the middle of the layer (not shown) whereas the skewness remains unchanged close to the boundaries.
The symmetry breaking between cyclonic and anti-cyclonic structures is therefore an underlying property of small-scale rotating Rayleigh-B\'enard convection and not the result of this large-scale circulation.
The skewness decreases as the Rossby number is decreased, which is consistent with the asymptotic model of Julien \textit{et al.}\cite{julien2012} and with simulations of decaying rotating turbulence\cite{bokhoven2008} and rotating Rayleigh-B\'enard convection experiments\cite{vorobieff2002,kunnen2010}.

Another interesting observation comes from the embedded plot in Figure~\ref{fig:pdfw} that focuses on small values of the vertical vorticity.
Recall that the vertical vorticity is scaled with the value of the background vorticity $2\Omega$.
For cyclonic structures, the critical value $\omega=2\Omega$ does not affect the vorticity distribution and we observed an exponential tail.
However, the PDF of anti-cyclonic vorticity drops for absolute values of the vertical vorticity that are larger than the background vorticity.
This indicates that anti-cyclonic structures are less likely to have large values of vertical vorticity than cyclonic structures.
This transition is particularly visible at low Rossby numbers and low Reynolds numbers where the flow is organised in vortical cells aligned with the rotation axis (see the top panel in Figure~\ref{fig:side}).
As the Rayleigh number is increased from one case to the next, this transition becomes gradually less clear.
We observed this sharp transition in the vertical vorticity PDF when $\omega_z\approx-2\Omega$ for all Taylor numbers considered. 
This transition in the behaviour of the negative vorticity is possibly related to instabilities observed in idealised vortices with a rotating background.
It is known that centrifugal instabilities are not possible for cyclonic vortices but are possible for anti-cyclonic vortices provided that that $W_0/2\Omega<-1$ where $W_0$ is the vorticity at the core of the vortex\cite{ooyama1966,sipp1999,godeferd2001}.
This seems to be consistent with the fact that value of the vertical vorticity smaller than $-2\Omega$ are very unlikely.
As the Rayleigh number is increased, the bulk of the flow becomes more turbulent and 3D so that this argument based on idealised vortices is less applicable, although a clear asymmetry is still observed.
Note also that Vorobieff \textit{et al}.\cite{vorobieff2002} found experimentally a correspondence between positive vertical vorticity and stable focus topology of the motion and between negative vertical vorticity and unstable focus topology. 

Irrespective of the existence of an inverse cascade and the associated large-scale depth-invariant horizontal flow, rotating Rayleigh-B\'enard flows with stress-free boundaries tends to favour strong cyclonic motions since the anti-cyclonic vortices close to the boundaries are unstable to centrifugal instabilities.
If the Rossby number is low enough and if the Reynolds number is large enough, this asymmetry in the vorticity distribution of the small-scale convective motions will persist in the large-scale circulation leading to a condensate state made of one large-scale coherent cyclonic vortex and spread out incoherent anti-cyclonic motions (see the bottom panels in Figure~\ref{fig:temp}).
In the asymptotic regime of low Rossby numbers, symmetry is restored and large-scale cyclones can coexist with large-scale anti-cyclones\cite{rubio2014}.
%
%
\section{Conclusion\label{sec:concl}}

We performed numerical simulations of rapidly-rotating Rayleigh-B\'enard convection using the Boussinesq approximation.
We have focused on the rapidly-rotating regime where it has been observed using differently constructed models\cite{chan2007,kapyla2011,mantere2011,julien1998,julien2012} that a large-scale horizontal circulation can be sustained by the small-scale turbulent convection.
We confirm that such a large-scale circulation is also observed in the Boussinesq limit using the full set of equations valid for any values of the Rossby number.

The existence of such a large-scale horizontal flow depends on two conditions.
First, the initial small-scale flow must be sufficiently turbulent.
For all parameters considered here, no large-scale circulation was observed very close to the onset of convection where the flow is organised in laminar and quasi-steady columnar structures aligned with the rotation axis.
The Reynolds number based on the correlation length-scale of the horizontal flow and the root mean square velocity (see equation \ref{eq:re}) must be approximately larger than 100 in order for the flow to generate a dominant vortex mode.
Second, the flow must be sufficiently constrained by rotation.
We quantify this constraint in terms of the Rossby number (as defined by equation \eqref{eq:ro}), which should be of order unity, or lower.
If the first condition is met (\textit{i.e.} the flow is turbulent) but the Rossby number is too large, no large-scale flow will be observed.
This is the case for example for Taylor numbers smaller than $Ta=10^8$, where we did not observe large-scale flows independently of the value taken for the Rayleigh number (see Figure~\ref{fig:param}).
We also observed the eventual disappearance of the large-scale circulation at very large Rayleigh number for a fixed Taylor number.
As the Rayleigh number is increased between different simulations, the flow undergoes a transition between strongly anisotropic turbulence (characterized by ``cigar''-shaped structures) to a quasi-isotropic state of turbulence (where the Reynolds stress tensor tends towards an isotropic state) in the middle of the layer.
Eventually, as the flow becomes more turbulent and isotropic, the Rossby number becomes too large so that no large-scale flow can be sustained.
We observed this restoration of isotropy for particular cases at $Ta=10^8$ and $Ta=10^9$ (cases $C5$, $C6$ and $D6$ in Table~\ref{tab:one}), but we speculate that this will be observed for any Taylor numbers and sufficiently large Rayleigh numbers.
Unfortunately, we could not check this due to the obvious numerical constraints at this present time, but our data suggest that this critical value of the Rayleigh number (above which the vortex mode is not sustained) scales approximately as the Taylor number.
Note that the scaling $Ra\approx Ta$ also corresponds to constant values for the Rossby number as defined by equation~\eqref{eq:ro}, which is consistent with our previous analysis. 
 
The origin of the large-scale flow is due to a non-local inverse cascade of energy and the 2D behaviour of the depth-averaged vortex mode, which was already observed in the asymptotic model of Julien \textit{et al.}\cite{julien2012}.
We have confirmed this observation using the full set of equations, and a different shell-by-shell energy transfer analysis.
Our results are very similar to energy transfers observed in forced rotating turbulence in tri-periodic domains\cite{mininni2009}, which confirms that this inverse cascade mechanism is a generic feature of rotating flows at low Rossby and larger Reynolds numbers.
As in 2D turbulence, we also observed vortex merging events, provided that the aspect ratio of the numerical domain is large enough.

The asymmetry between cyclonic and anti-cyclonic vortices is not related to the existence of this large-scale circulation.
It is an underlying property of small-scale 3D wave mode generated by rapidly-rotating convection.
We showed that the vertical vorticity statistics barely evolve as the large-scale flow is growing in amplitude.
From the beginning, a clear asymmetry is observed between different sign of the vertical vorticity and cyclonic structures are dominant, especially close to the boundaries, where the flow can be described as a collection of localised 2D vortices aligned with the rotation axis.
A clear transition is observed when the negative anti-cyclonic vertical vorticity close to the boundaries equals the background vorticity $2\Omega$.
Large values of negative vertical vorticity are less likely than values smaller than the background vorticity whereas the cyclonic vorticity has a long exponential tail without particular transitions.
As observed in stability analyses of idealised vortex with a background rotation\cite{godeferd2001}, anti-cyclonic vortices can be destabilised by centrifugal instabilities whereas cyclonic structures remain stable.

Several simplifications were made in this work and it is important to discuss these here.
First, we only considered stress-free boundary conditions.
As discussed in section \ref{sec:cycl}, no-slip boundary conditions are also of interest since the Ekman pumping at the boundary might significantly change the dynamics of vorticity and large-scale horizontal flow, but these are beyond the remit of this paper.
Imposing a flux instead of fixing the temperature at the boundaries might also affect the structure of the large-scale flow since it is associated with a decrease in the Nusselt number\cite{guervilly2014}.
Second, we only considered the incompressible limit in the Boussinesq approximation.
The most striking difference between our incompressible simulations and previously published compressible models is the ability of compressible rotating convection to sustain large-scale anti-cyclonic structures\cite{mantere2011}.
As already mentioned, we only observed cyclonic large-scale vortices in our moderate Rossby numbers simulations.
Since the horizontally-averaged vertical vorticity vanishes due to our choice of boundary conditions, anti-cyclonic vorticity must exist but is weak and spread amongst the layer.
The asymptotic model of Julien \textit{et al.}\cite{julien2012} showed that in the vanishing Rossby limit, a symmetric state of geostrophic turbulence populated with both cyclones and anti-cyclones is obtained.
We did not observe such a symmetric state due to the moderate values of the Rossby number considered here ($Ro_L>10^{-2}$, see Table~\ref{tab:one}), or a dominant large-scale anti-cyclone as in the compressible regime.
The possibility of coherent anti-cyclonic large-scale vortices seems to be a specific of the compressible model.
We are investigating this aspect of the problem, and results on how density stratification drastically changes the angular momentum redistribution in the flow will be discussed in a subsequent paper.

Simulations have been performed on the HPC facility Archer under the Project Group e308.
The authors would like to thank C. Guervilly and P.J. Bushby for useful numerical comparisons and discussions.

\bibliographystyle{ieeetr}
\bibliography{biblio}

\end{document}